\documentclass[aps,prd,preprintnumbers,reprint,superscriptaddress]{revtex4-1}

\usepackage{amsmath}
\usepackage{amssymb}
\usepackage{amsthm}
\usepackage{mathtools}
\usepackage{mathrsfs}
\usepackage{bm}
\usepackage{slashed} 
\usepackage{graphicx}
\usepackage{multirow}
\usepackage{tikz}
\usepackage[caption=false]{subfig}
\usepackage{relsize}	
\usepackage{array}
\usepackage{float}
\usepackage{color}
\usepackage{xcolor}
\usepackage{soul}
\usepackage{verbatim} 
\usepackage{siunitx}
\usepackage{hyperref}
\hypersetup{colorlinks=true,linkcolor=purple,anchorcolor=blue,citecolor=purple1, filecolor=blue,urlcolor=red,bookmarksnumbered=true,
pdfview=FitB
}
\allowdisplaybreaks
\usepackage{cleveref}

\def\be{\begin{eqnarray}}
\def\ee{\end{eqnarray}}
\colorlet{purple1}{blue!70!red}
\colorlet{darkred}{red!50!black}

\def\dd{{\mathrm{d}}}

\mathchardef\-="2D

\begin{document}
\title{Parton distribution functions of heavy mesons on the light front}

\author{Jiangshan Lan}
\email{jiangshanlan@impcas.ac.cn}
\affiliation{Institute of Modern Physics, Chinese Academy of Sciences, Lanzhou 730000, China}
\affiliation{School of Nuclear Science and Technology, University of Chinese Academy of Sciences, Beijing 100049, China}
\affiliation{Lanzhou University, Lanzhou 730000, China}
\affiliation{CAS Key Laboratory of High Precision Nuclear Spectroscpy, Institute of Modern Physics, Chinese Academy of Sciences, Lanzhou 730000, China}

\author{Chandan Mondal}
\email{mondal@impcas.ac.cn} \affiliation{Institute of Modern Physics, Chinese Academy of Sciences, Lanzhou 730000, China}
\affiliation{School of Nuclear Science and Technology, University of Chinese Academy of Sciences, Beijing 100049, China}
\affiliation{CAS Key Laboratory of High Precision Nuclear Spectroscpy, Institute of Modern Physics, Chinese Academy of Sciences, Lanzhou 730000, China}

\author{Meijian Li}
\email{meijianl@iastate.edu} \affiliation{Department of Physics and Astronomy, Iowa State University, Ames, Iowa 50011, USA}
\affiliation{Department of Physics,  University of Jyv\"{a}skyl\"{a}, P.O. Box 35, FI-40014, Jyv\"{a}skyl\"{a}, Finland}
\affiliation{Helsinki Institute of Physics, University of Helsinki, P.O. Box 64, FI-00014, Helsinki, Finland}

\author{Yang Li}
\email{leeyoung@iastate.edu}
\affiliation{School of Nuclear Science and Technology, University of Chinese Academy of Sciences, Beijing 100049, China}
 \affiliation{Department of Physics and Astronomy, Iowa State University, Ames, Iowa 50011, USA}


\author{Shuo Tang}
\email{tang@iastate.edu} \affiliation{Department of Physics and Astronomy, Iowa State University, Ames, Iowa 50011, USA}

\author{Xingbo Zhao}
\email{xbzhao@impcas.ac.cn} \affiliation{Institute of Modern Physics, Chinese Academy of Sciences, Lanzhou 730000, China}
\affiliation{School of Nuclear Science and Technology, University of Chinese Academy of Sciences, Beijing 100049, China}
\affiliation{CAS Key Laboratory of High Precision Nuclear Spectroscpy, Institute of Modern Physics, Chinese Academy of Sciences, Lanzhou 730000, China}

\author{James P. Vary}
\email{jvary@iastate.edu} \affiliation{Department of Physics and Astronomy, Iowa State University, Ames, Iowa 50011, USA}
\collaboration{BLFQ Collaboration}

\begin{abstract}
The parton distribution functions (PDFs) of heavy mesons are evaluated from their light-front wave functions, which are obtained from a basis light-front quantization in the leading Fock sector representation. We consider the mass eigenstates from an effective Hamiltonian consisting of the confining potential adopted from light-front holography in the transverse direction, a longitudinal confinement, and a one-gluon exchange interaction with running coupling. We present the gluon and the sea quark PDFs which we generate dynamically from the QCD evolution of the valence quark distributions.
\end{abstract}


\maketitle
\section{Introduction}
Heavy quarkonium is a multiscale system with all regimes of quantum chromodynamics (QCD). The perturbative expansion in the strong coupling constant $\alpha_s(\mu^2)$ is possible at high energies; however, at low energies, nonperturbative effects dominate. Heavy quarkonium provides an ideal platform for testing the interplay between perturbative and nonperturbative QCD within a data-rich regime. Production of heavy quarkonium, i.e. charmonium ($c\bar{c}$) and bottomonium ($b\bar{b}$), takes place via initial partonic scattering processes with large momentum transfer on a time scale of $\hbar/(2 {\rm m}_{\rm{[c,b]}}c^2)$, where $\rm m$ is the mass of the quark \cite{Acharya:2018upq}. Enormous progress has been made on $c\bar{c}$ and $b\bar{b}$ decays, showing that many measurements of branching fraction, width, and spectra have attained high precision (see Ref. \cite{Brambilla:2010cs} and the references therein). However, data on the $B_c$ meson ($b\bar c$ or $c \bar b$) family, which is unique since they are composed of two flavors of heavy quark and antiquark, are relatively scarce. So far, the ground state and its first radial excitation are confirmed in experiments \cite{Abe:1998fb,Aad:2014laa}. Meanwhile, creations of a large ensemble of heavy mesons are expected from ongoing and forthcoming high energy experiments, e.g. Large Hadron Collider (LHC) and Relativistic Heavy Ion Collider (RHIC), attracting dedicated theoretical efforts for understanding their structure  \cite{Koponen:2017fvm,Colquhoun:2015oha,Ali:2016gdg,Bhattacharya:2017aao,Li:2015zda,Jia:2015pxx,Li:2017mlw,Tang:2018myz,Tang:2019gvn}. 
	
In contrast to light mesons, heavy quarkonia are arguably among the simplest mesons. The constituent quark and antiquark are quite heavy  and move rather slowly inside the meson bound states. These two essential features ensure the hierarchical structure of the intrinsic energy scales of a quarkonium. The influential nonrelativistic QCD (NRQCD) factorization approach \cite{Bodwin:1994jh} fully employs this scale hierarchy and allows us to efficiently separate the relativistic and perturbative contributions from the long-distance and nonperturbative dynamics. Unlike the parton distribution amplitudes (PDAs) for light mesons which are totally nonperturbative objects, the PDA for heavy mesons can be factorized into a product of a perturbatively calculable distribution part and a NRQCD matrix-element for the vacuum to hadron state transition at the lowest order in velocity expansion \cite{Ma:2006hc,Bell:2008er}. The profile of the quarkonium PDA is fully acquiescent to perturbation theory. Although the heavy quarkonium PDAs have become objects of intensive study \cite{Jia:2015pxx,Li:2017mlw,Tang:2018myz,Ma:2006hc,Bell:2008er,Bodwin:2006dm,Braguta:2006wr,Braguta:2007fh,Braguta:2007tq,Choi:2007ze,Braguta:2008qe,Hwang:2008qi,Hwang:2009cu,Xu:2016dgp,Wang:2017bgv,Hwang:2010hw}, there is limited knowledge of parton distribution functions (PDFs) of heavy mesons. 

PDFs, appearing in the description of hard inclusive reactions like deep inelastic scattering (DIS), play important roles in  understanding the structure  of hadrons. PDFs encode the distribution of longitudinal momentum and polarization carried by the constituents. 
There are many experiments and theoretical investigations on this subject and it remains an active field of research over many years. For example, the measurement of the inclusive charm ($c$) and bottom ($b$) quark cross sections in DIS at DESY-HERA uniquely constrains the PDFs of the proton, in particular, its $b$ and $c$ content \cite{Aaron:2009af}. The predictions of the inclusive production of $W$ and $Z$ bosons, are sensitive to the theoretical treatment of heavy quarks \cite{Collins:1998rz,Aivazis:1993kh,Aivazis:1993pi,Kramer:2000hn,Tung:2001mv,Tung:2006tb,Thorne:1997ga,Thorne:1997uu,Thorne:2006qt,Buza:1997nv,Chuvakin:1999nx,Chuvakin:2000jm,Kretzer:2003it,Martin:2007bv,Alekhin:2008hc,Bierenbaum:2009zt,Bierenbaum:2009mv,Nadolsky:2009ge,Thorne:2008xf,Nadolsky:2008zw,Martin:2009iq}. The bottom quark PDF is crucial in Higgs production at the LHC in both the Standard Model and in extensions to the Standard Model \cite{Dicus:1998hs,Huang:1998vu,Balazs:1998sb,Campbell:2002zm,Maltoni:2003pn}. The PDFs of heavy quarks within the nucleon have been extensively investigated, however, little is known from either theory or experiment about the PDFs of the heavy mesons, although this situation is likely to change with the new LHC and RHIC programs on heavy mesons.


In this paper, we evaluate the unpolarized PDFs of heavy quarkonia and $B_c$ mesons using the light-front wave functions (LFWFs) based on a basis light-front quantization (BLFQ) approach \cite{Vary:2009gt} where only the leading Fock sector has been considered. In the effective Hamiltonian, we choose the confining potential adopted from the light-front holography in the transverse direction \cite{Brodsky:2014yha}, a longitudinal confinement \cite{Li:2015zda}, and a one-gluon exchange interaction with a running coupling. The nonperturbative solutions for the LFWFs are provided by the recent BLFQ study  of heavy quarkonia \cite{Li:2017mlw} and $B_c$ mesons \cite{Tang:2018myz}. The LFWFs have been successfully applied to compute the decay constants, r.m.s. radii, distribution amplitudes, electromagnetic form factor etc. of heavy mesons~\cite{Li:2017mlw,Tang:2018myz}. We extend our investigations to study the QCD evolution of the heavy meson PDFs in order to obtain the gluon and the sea quark distributions. Most of the gluons and sea quarks are expected to be produced dynamically by the scale evolution. Here, we consider only the leading Fock sectors of the Fock state expansion for the quarkonia and $B_c$ meson states. We use the Dokshitzer-Gribov-Lipatov-Altarelli-Parisi (DGLAP) equation of QCD \cite{Dokshitzer:1977sg,Gribov:1972ri,Altarelli:1977zs} up to the next-to-next-to-leading order (NNLO)  for the evolution of the valence quark PDFs and obtain the gluon and the sea quark PDFs. Since, the DGLAP evolution is applicable in the perturbative regime, the large mass scales of heavy mesons justify the use of the QCD evolution. Studying the DGLAP evolution of heavy quark PDFs  provides rich information about the gluon and sea quark appearing in higher Fock sectors (such as $q \bar q g$ and $q \bar q \bar q q $). Our study thereby provides guidance for the structure of heavy mesons at higher scales.

The paper is organized as follows. We discuss the BLFQ formalism for heavy meson systems in Sec.~\ref{BLFQ}. The PDFs of heavy mesons have been evaluated in Sec.~\ref{pdf}. The scale evolution of the heavy quarkonium and $B_c$ meson PDFs has also been discussed in this section. We summarize in Sec.~\ref{summary}.
\section{Basis Light-Front Quantization \label{BLFQ}}
BLFQ approach is developed for solving many-body bound-state problems in quantum field theory \cite{Vary:2009gt,Wiecki:2014ola,Li:2015zda}. It is a Hamiltonian-based formalism which takes advantage of light-front dynamics \cite{Brodsky:1997de}. This approach has been successfully applied to quantum electrodynamics (QED) systems such as the single electron problem~\cite{Zhao:2014xaa}, as well as the strong coupling bound-state positronium problem \cite{Wiecki:2014ola} and QCD systems such as the running coupling quarkonium problem~\cite{Li:2017mlw}. It has also been applied to the $B_c$ mesons~\cite{Tang:2018myz}. Recently, the BLFQ approach using a Hamiltonian that includes the color singlet Nambu--Jona-Lasinio interaction to account for the chiral dynamics has been applied to the light mesons \cite{Jia:2018ary,Lan:2019vui,Lan:2019rba}. Furthermore, the BLFQ formalism has been extended to time-dependent strong external field problems such as nonlinear Compton scattering~\cite{Zhao:2013jia}. (For the reviews related to BLFQ and its other application, see Refs.~\cite{Leitao:2017esb,Li:2018uif,Adhikari:2016idg,Chen:2016dlk,Li:2017uug,Chakrabarti:2014cwa,Adhikari:2018umb,Du:2019ips,Mondal:2019zzz}.)

The effective light-front Hamiltonian for the heavy meson consists of the light-front kinetic energy with a harmonic oscillator confining potential in the transverse direction, based on the light-front holography, 
 as well as a longitudinal confining potential, and the one-gluon exchange interaction with a running coupling. In a light-front Hamiltonian approach~\cite{Vary:2009gt}, a recent study of heavy meson presents the effective Hamiltonian as \cite{Li:2015zda,Li:2017mlw,Tang:2018myz},
\be\label{eqn:Heff}
&&  H_\mathrm{eff} = T_q+T_{\bar{q}}+V_{\rm conf}+V_{\rm OGE},
\ee
where $T_{q(\bar{q})}$ is the kinetic energy of the quark (antiquark). $V_{\rm conf}$ represents the confining potential which includes both the transverse and the longitudinal confinements and $V_{\rm OGE}$ is the one-gluon exchange term. 
In the second quantized form, the effective Hamiltonian 
 can be written in momentum-space variables as,
\begin{widetext}
\begin{equation}\label{eq2}
\begin{split}
H_\textrm{eff} =\,& \frac{1}{2}\sum_{\lambda_{ q},\lambda_{\bar q}} \int \frac{\dd^3P}{2(2\pi)^3P^+} \frac{\dd x}{2x(1-x)} \frac{\dd^2\vec{k}_\perp}{(2\pi)^3} 
\Big[ \frac{\vec k_{\perp}^2+m^2_q}{x} + \frac{\vec {k}_{\perp}^2+m_{\bar q}^2}{1-x} \Big] \cdot b^\dagger_{\lambda_{q}}(p_1) d^\dagger_{\lambda_{\bar q}}(p_2) d_{\lambda_{\bar q}}(p_2)b_{\lambda_{q}}(p_1) \\
-  \,&\sum_{\lambda_q, \lambda_q ', \lambda_{\bar q}, \lambda_{\bar q}'} \int \frac{\dd^3P}{2(2\pi)^3P^+} \frac{\dd x}{2x(1-x)} \frac{\dd^2\vec{k}_\perp}{(2\pi)^3} 
 \frac{\dd x'}{2x'(1-x')} \frac{\dd^2\vec{k}'_\perp}{(2\pi)^3} 
  \Big[ \\
\times  \,& 2(2\pi)^3\kappa^4 xx'(1-x)(1-x') [\nabla^2_{k_\perp} \delta^2(\vec k_\perp-\vec k'_\perp)]\delta(x-x') \delta_{\lambda_q\lambda_{q'}} \delta_{\lambda_{\bar q}\lambda_{{\bar q}'}}\\
  + \,& 
   2(2\pi)^3x'(1-x')\frac{\kappa^4}{(m_q+m_{\bar q})^2} \partial_x\big(x(1-x)\partial_x\delta(x-x')\big) \delta^2(\vec k_\perp-\vec k'_\perp)\delta_{\lambda_q\lambda_{q'}} \delta_{\lambda_{\bar q}\lambda_{{\bar q}'}} \\
  + \,& \frac{C_F 4\pi\alpha_s(Q^2)}{Q^2} \bar u_{\lambda_{\bar q}'}(p_1')\gamma_\mu u_{\lambda_{q}}(p_1) \bar v_{\lambda_{\bar q}}(p_2)\gamma^\mu v_{\lambda_{\bar q}'}(p_2')
  \Big]  
  \cdot b^\dagger_{\lambda_q'}(p_1')d^\dagger_{\lambda_{\bar q}'}(p_2') d_{\lambda_{\bar q}}(p_2) b_{\lambda_{q}}(p_1), \\
\end{split}
\end{equation}
\end{widetext}
where the momenta of quark and antiquark are $p_1 \equiv (p_1^-,p_1^+,\vec{p}_{1\perp})= \big( \frac{(\vec k_\perp+x\vec P_\perp)^2+m_q^2}{xP^+}, xP^+, \vec k_\perp+x\vec P_\perp \big)$ and
$p_2 \equiv (p_2^-,p_2^+,\vec{p}_{2\perp})= \big( \frac{(-\vec k_\perp+(1-x)\vec P_\perp)^2+m_{\bar q}^2}{(1-x)P^+}, (1-x)P^+, -\vec k_\perp+(1-x)\vec P_\perp \big)$, respectively. The definitions of $p_1'$ and $p_2'$ are similar.  
We are working with color singlet states and we have suppressed  the color indices.
Here, ${\rm m}_q$ (${\rm m}_{\bar q}$) is the mass of the quark (antiquark), and $\kappa$ is the strength of the confinement. $\partial_x\equiv (\partial/\partial x)_{\vec \zeta_\perp}$, where
$\vec \zeta_\perp \equiv \sqrt{x(1-x)} \vec r_\perp$ is the holographic variable \cite{Brodsky:2014yha}. $\vec r_\perp$ measures the transverse separation between the quark and the antiquark. In momentum space, $\langle \vec p_\perp|\widehat {r^2}_\perp|\vec p_\perp '\rangle=-\nabla^2_{p_\perp}\delta^2(\vec p_\perp-\vec p_\perp ')$.
$C_F = (N_c^2-1)/(2N_c)=4/3$ is the color factor for the color singlet state and 
$Q^2 
 = \frac{1}{2}\Big(\sqrt{\frac{x'}{x}}\vec {k}_\perp-\sqrt{\frac{x}{x'}}\vec {k'}_\perp\Big)^2 +
\frac{1}{2}\Big(\sqrt{\frac{1-x'}{1-x}}\vec {k}_\perp-\sqrt{\frac{1-x}{1-x'}}\vec {k'}_\perp \Big)^2 
+ \frac{1}{2}(x-x')^2\Big( \frac{m_q^2}{xx'}+\frac{m_{\bar q}^2}{(1-x)(1-x')}\Big) + \mu_g^2,$ 
 is the average 4-momentum squared carried by the exchanged gluon. We employ the running coupling $\alpha_s(Q^2)$ based on the 1-loop perturbative QCD. A finite vector boson mass $\mu_g=0.02$ GeV has been introduced to regularize the integrable Coulomb singularity~\cite{Li:2017mlw}. The $Q^2$ in the denominator of the boson exchange interaction arises from the cancellation of the light-front small-$x$ divergences which appear in the instantaneous vector boson-exchange interaction. A detailed discussion about  the cancellation of the instantaneous interaction in the effective vector-boson exchange interaction can be found in Ref.~\cite{Wiecki:2014ola}.
 
In light-front holography, the confinement $\kappa^4\zeta_\perp^2$ is introduced in the massless case. For heavy mesons, the quark masses and the longitudinal dynamics cannot be neglected and thus, a longitudinal confining potential (third term in Eq.~(\ref{eq2})~\cite{Li:2015zda}) has been introduced to complement the transverse holographic confinement. We combine the holographic potential and one-gluon exchange interaction to govern  the long distance as well as the short distance physics. 
 Here, we retain only contributions to the effective Hamiltonian which are relevant to the simplest Fock space. 
Note that the Dirac matrix structure of the two confining potentials is the identity matrix.
	However, we manifest them only with the  good (independent) component of the quark fields. The Hamiltonian is expressed only in terms of the independent degrees of freedom which are conventionally selected to be the Pauli spinors. 
	The dependent degrees of freedom are accessible through the equations of motion \cite{Wiecki:2014ola}.
	
The quark and antiquark creation (annihilation) operators $b^\dagger$ and $d^\dagger$ ($b$ and $d$) satisfy the following canonical anticommutation relations,
\be\label{eqn:CCR}
&&\quad \big\{ b_{{\lambda_{ q}}i}(p^+, \vec p_\perp), b_{{\lambda'_{  q}}i'}^\dagger(p'^+, \vec p'_\perp) \big\} \nonumber\\
&& = \big\{ d_{{\lambda_{ q}}i}(p^+, \vec p_\perp), d_{{\lambda'_{  q}}i'}^\dagger(p'^+, \vec p'_\perp) \big\} \nonumber\\
&& = 2p^+(2\pi)^3\delta^3(p-p')\delta_{{{\lambda_{ q}}{\lambda'_{q}}}}\delta_{ii'},
\ee
where $\delta^3(p-p') \equiv \delta(p^+-p'^+)\delta^2(\vec p_\perp-\vec p'_\perp)$.

The spectrum and light-front Fock state wave functions are obtained from the solution of the mass (${\rm M}$) eigenvalue equation 
\be H_{\text{eff}}|\psi^J_{m_J}\rangle= {\rm M}^2|\psi^J_{m_J}\rangle, \label{eq:Heff_eigen_value} 
\ee
where the Fock space representation of the heavy meson state $|\psi^J_{m_J}\rangle$ reads:
\be\label{eqn:Fock_expansion}
 |\psi^J_{m_J}\rangle &=& 
\sum_{\lambda_{q},\lambda_{\bar q}}\int_0^1\frac{d x}{2x(1-x)} \int \frac{d^2 \vec k_\perp}{(2\pi)^3}
\, \psi^{(m_J)}_{\lambda_{q},\lambda_{\bar q}}(\vec k_\perp, x)\nonumber\\
&&\times\frac{1}{\sqrt{N_c}}\sum_{i=1}^{N_c} b^\dagger_{\lambda_{ q}{}i}\big(xP^+, \vec k_\perp+x\vec P_\perp\big)\nonumber\\
&&\times d^\dagger_{\lambda_{\bar q}{}i}\big((1-x)P^+,
-\vec k_\perp+(1-x)\vec P_\perp\big) |0\rangle. 
\ee
The coefficients of the expansion, $\psi^{(m_J)}_{{\lambda_{ q}}{\lambda_{\bar q}}}(\vec k_\perp, x)$, are the valence sector LFWFs with $\lambda_{ q}$ ($\lambda_{\bar q}$) representing the spin of the quark (antiquark), $i$ is the color index of the quark (antiquark). The superscript ``$m_J$" signifies we are working on a basis with fixed total angular momentum projection. We will henceforth suppress this superscript. 

To evaluate the Hamiltonian matrix, one needs to construct the basis. In order to construct the basis, the two-dimensional (2D) harmonic oscillator (HO) functions are adopted in the transverse direction, which are defined in terms of the dimensionless transverse momentum variable 
(${\vec q}_\perp/b$) as~\cite{Li:2015zda}:
\be
 \phi_{nm}(\vec q_\perp; b) =&& b^{-1} \sqrt{\frac{4\pi n!}{(n+|m|)!}} \bigg(\frac{q_\perp}{b}\bigg)^{|m|}\exp\big(-q^2_\perp/(2b^2)\big)\nonumber\\
&&\times  L_n^{|m|}(q^2_\perp/b^2) \exp\big(i m \theta_q),  \label{eq:transverse_basis}
\ee
 where $\vec q_\perp \triangleq \vec k_\perp/\sqrt{x(1-x)}$, $q_\perp = |\vec q_\perp|$, $\theta_q = \arg \vec q_\perp$, $b$ is the HO basis scale parameter with dimension of mass, $n$ and $m$ are the radial and the angular quantum numbers, $L_n^{|m|}(z)$ is the associated Laguerre polynomial. In the longitudinal direction, the basis functions are defined as 
  \be
 \chi_l(x) =&& \sqrt{4\pi(2l+\alpha +\beta +1)} \sqrt{\frac{\Gamma(l+1) \Gamma(l+\alpha +\beta +1)}{\Gamma(l+\alpha +1) \Gamma(l+\beta +1)}}\nonumber\\
&& \times x^{\frac{\beta}{2}} (1-x)^{\frac{\alpha}{2}}P_l^{(\alpha, \beta)}(2x-1),\label{eq:longitudinal_basis}
 \ee
 where $P_l^{(\alpha, \beta)}(z)$ is the Jacobi polynomial, $\alpha =2{\rm m}_{\bar{q}}({\rm m}_q+{\rm m}_{\bar{q}})/\kappa^2$ and $\beta =2{\rm m}_q({\rm m}_q+{\rm m}_{\bar{q}})/\kappa^2$  are dimensionless basis parameters, and $l=0,~1,~2,...$. Using the basis functions given in Eqs.~\eqref{eq:transverse_basis} and~\eqref{eq:longitudinal_basis}, the expansion of momentum-space LFWFs can be expressed as \cite{Li:2015zda,Li:2017mlw,Yang2019},
\be\label{eqn:basis_representation}
 \psi_{\lambda_q\lambda_{\bar q}}(\vec k_\perp, x) =&& \sum_{n, m, l}  \langle n, m, l, \lambda_q, \lambda_{\bar q} | \psi^J_{m_J}\rangle  \nonumber\\
&& \times\phi_{nm}(\vec k_\perp/\sqrt{x(1-x)} ; b) \chi_l(x),
\ee
where the coefficients $\langle n, m, l,\lambda_q,\lambda_{\bar q}|\psi^J_{m_J}\rangle$ are obtained in the BLFQ basis space by diagonalizing the truncated Hamiltonian matrix. In order to numerically diagonalize $H_{\text{eff}}$, the infinite dimensional basis must be truncated to a finite dimension. Here, we apply the following truncation to restrict the quantum numbers \cite{Li:2015zda,Li:2017mlw},
\begin{equation}
 2n + |m| + 1 \le N_\mathrm{max}, \quad 0 \le l \le L_{\mathrm{max}},\label{eq:truncation1}
\end{equation}
where $L_{\text{max}}$ is the basis resolution in the longitudinal direction whereas $N_{\text{max}}$ controls the transverse momentum covered by 2D-HO functions. The $N_{\max}$-truncation gives a natural pair of ultraviolet (UV) and infrared (IR) cutoffs: $\Lambda_\textsc{uv} \simeq b\sqrt{N_{\max}}$,
$\lambda_\textsc{ir} \simeq b/\sqrt{N_{\max}}$, where $b=\kappa$ is the energy scale parameter of the oscillator basis. The total angular momentum $J$ is an approximate quantum number due to the breaking of the rotational symmetry by the Fock sector truncation and the basis truncation in the BLFQ approach.  However, the projection of the total angular momentum $(m_J)$ for the system is always conserved,
\be  m_J = m + \lambda_{q}+\lambda_{\bar{q}}
\label{eq:truncation2}.
\ee  
For fixed $N_{\rm max}$ and
$L_{\rm max}$ , the model parameters are fixed by fitting the experimental data of the mass eigenvalues in the $m_J=0$ sector \cite{Li:2017mlw, Tang:2018myz}. The model parameters are summarized in Table \ref{T12}.

\begin{table}
  \centering
  \caption{List of the model parameters \cite{Li:2017mlw,Tang:2018myz}. $\alpha_s(Q^2)$ is the running coupling with the flavor number, $N_f$. $\kappa$ and ${\rm m}_q$ are the confining strength and mass of the heavy quark, respectively. $N_{\rm max}$ and $L_{\rm max}$ are the truncation parameters in the transverse and the longitudinal direction, respectively.}
  \begin{tabular}{cccccc}
\\
  \hline   \hline 
         & ~$\alpha_s(0)$~&~$N_{f}$~&~$\kappa$ (GeV)~&~${\rm m}_q$ (GeV)~&~$N_{\rm max}=L_{\rm max}$\\ \hline           
               $c\bar{c}$ & 0.6 &4 &0.985&1.570&8  \\
               $b\bar{b}$  & 0.6 &5 &1.389&4.902&32  \\
               	$b\bar{c}$ & 0.6 &4 &1.196&4.902,~1.603&32  \\

   \hline \hline
  \end{tabular}
  \label{T12}
\end{table}

\begin{table}
  \centering
  \caption{Initial scale ($\mu_0$) of charmonium, bottomonium and $B_c$ meson PDFs. Three different values of $\mu_0$ are considered for the PDFs. ${\rm m}_q$ is the mass scale of heavy quark. The BLFQ results with basis truncation $N_{\rm max}$  correspond to the UV cutoffs $\mu_h$.}
  \begin{tabular}{cccc}
  \hline 
  \hline $\mu_0$ [GeV]&  Charmonium&Bottomonium&$B_c$ meson\\
         &~ $N_{\rm max}=8$~&~$N_{\rm max}=32$~&~$N_{\rm max}=32$\\
             \hline
              ${\rm m}_q$ &1.570  &4.902&4.902  \\
              $\mu_h$ \cite{Li:2017mlw,Tang:2018myz} & 2.80 &7.90 &6.77 \\
              $2\mu_h$ & 5.60 &15.80 &13.54 \\

  \hline \hline
  \end{tabular}
  \label{T1}
\end{table}

 \begin{figure}
 \subfloat[charmonia  \label{fig:inicc8}]{
 \includegraphics[width=0.44\textwidth]{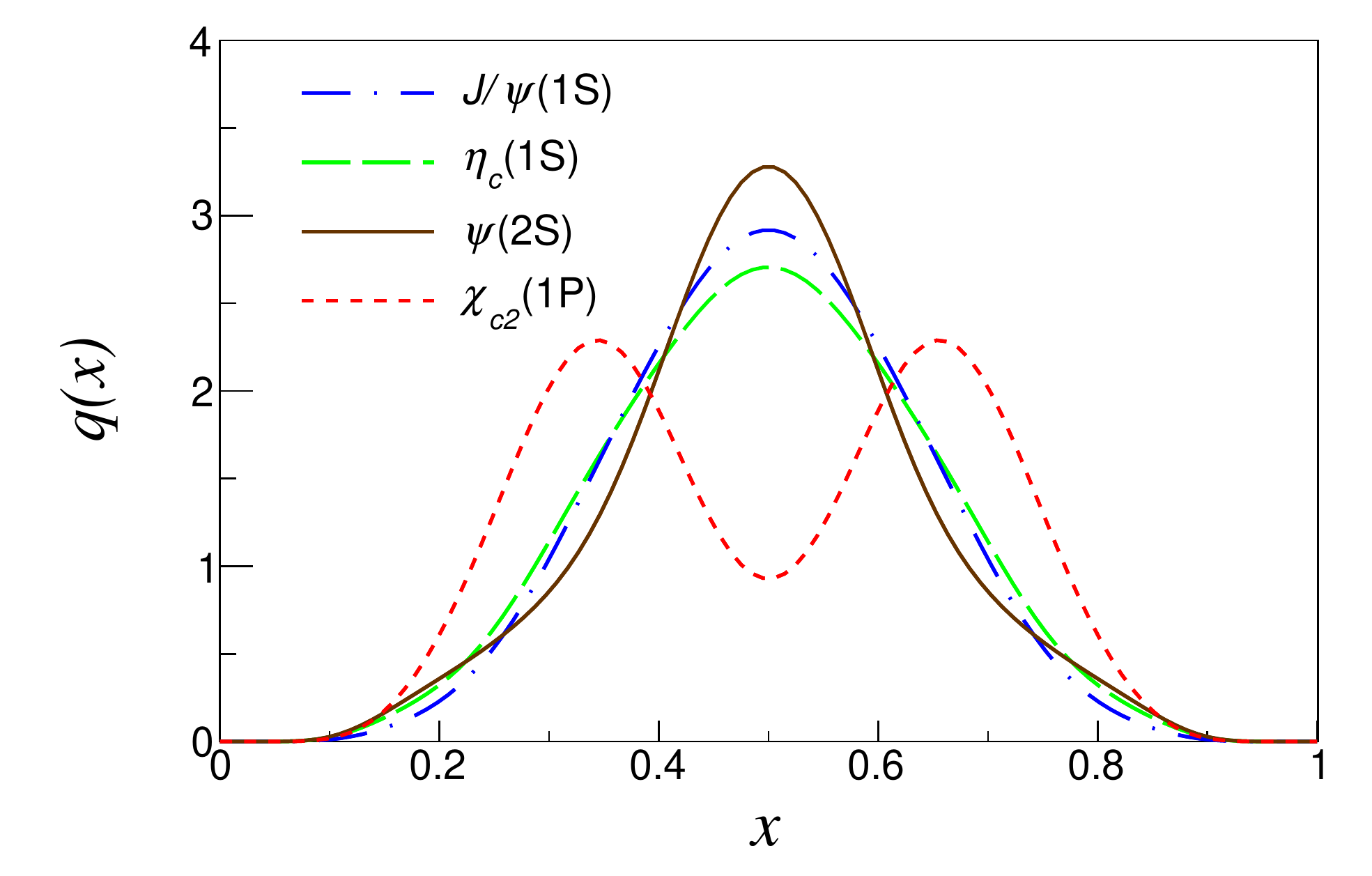}
}

 \subfloat[bottomonia  \label{fig:inicc32}]{
 \includegraphics[width=0.44\textwidth]{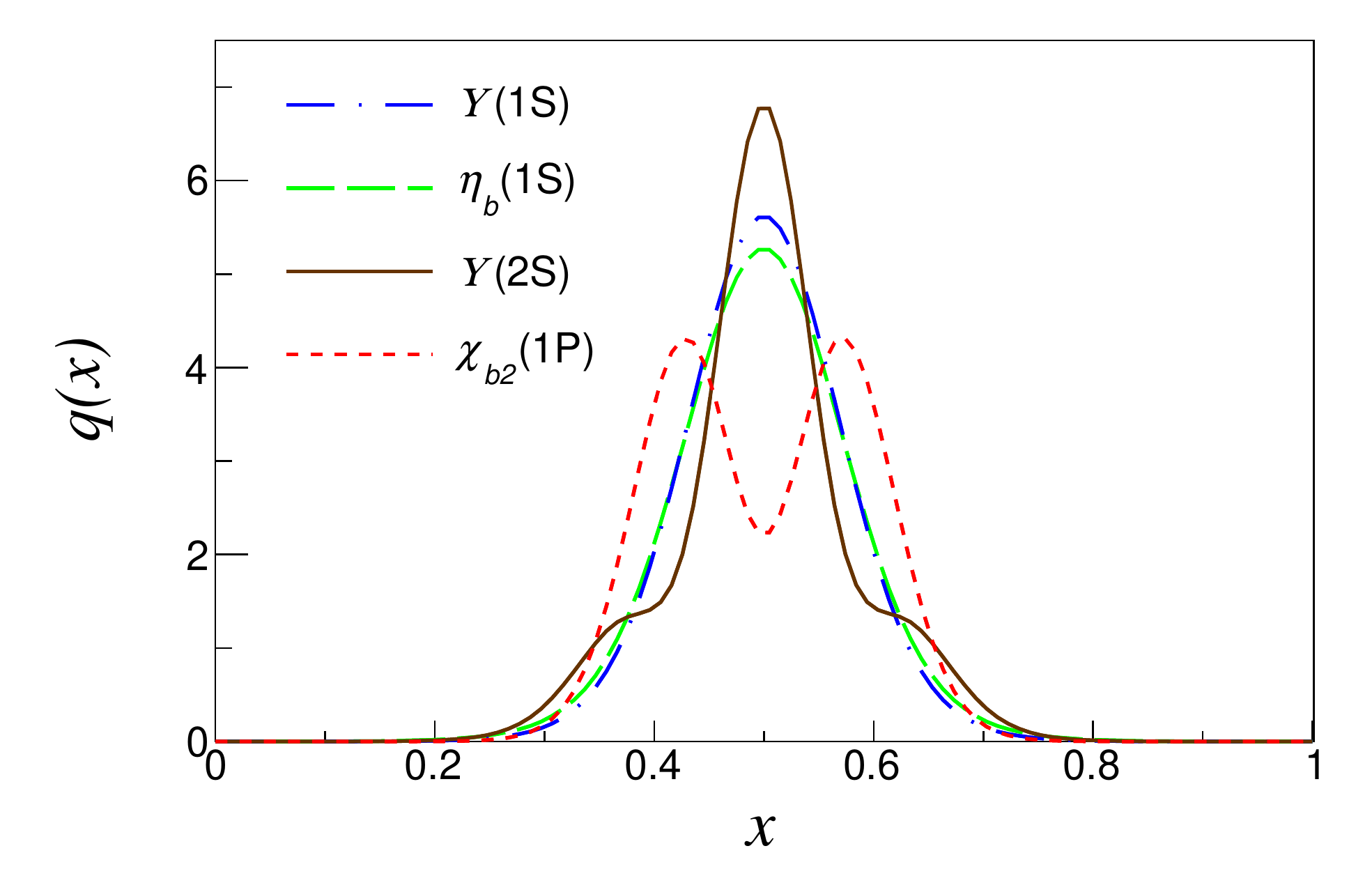}
}

  \subfloat[$B_c$ mesons  \label{fig:inibb8}]{
 \includegraphics[width=0.44\textwidth]{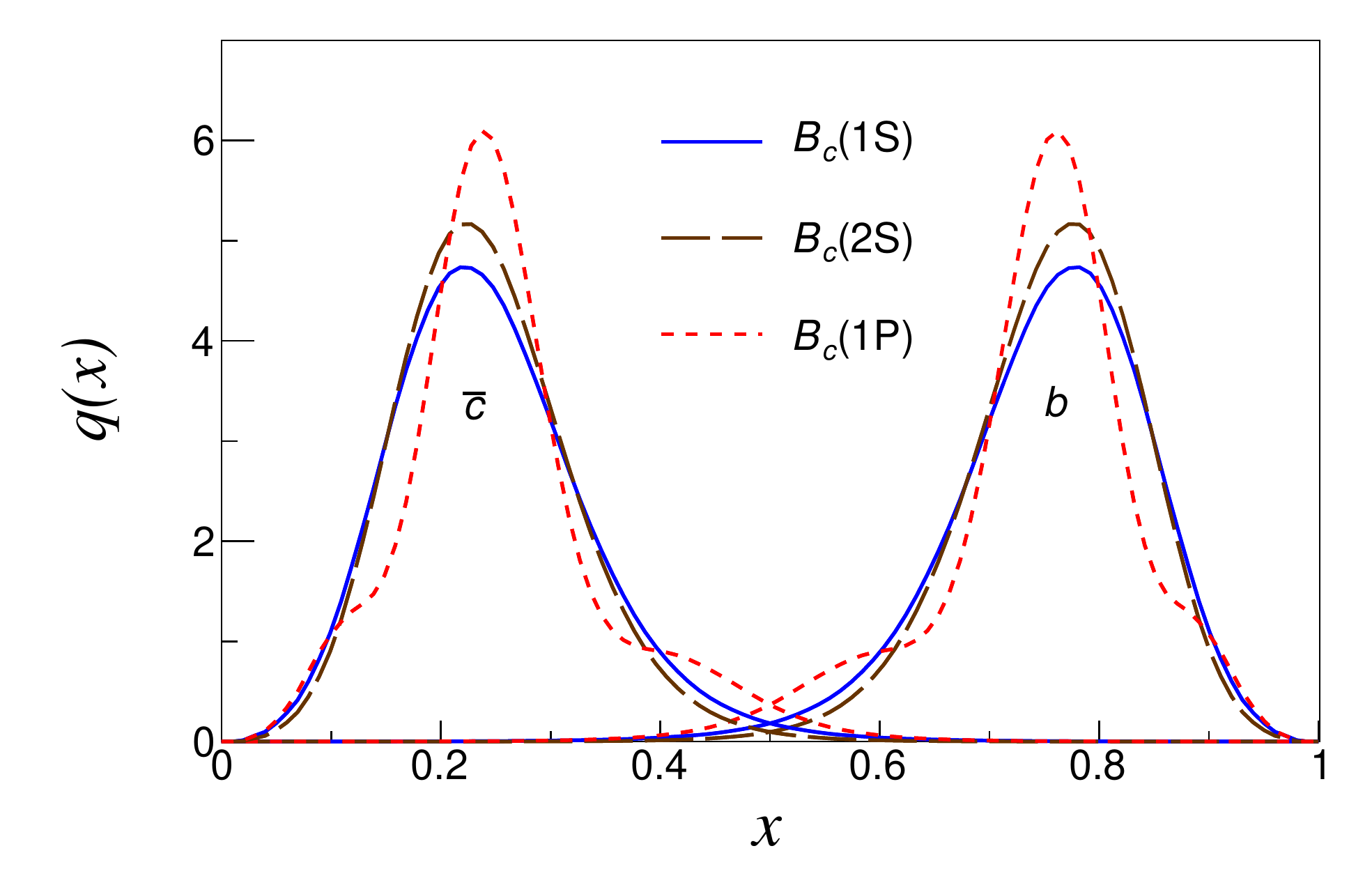}
}
\caption{PDFs of (a) $\eta_c(\rm 1S)$,~$J/\psi(\rm 1S)$,~ $\psi(\rm 2S)$,~ $\chi_{c2}(\rm 1P)$ (charmonium); (b) $\eta_b(\rm 1S)$, ~$\Upsilon(\rm 1S)$,~ $\Upsilon(\rm 2S)$,~ $\chi_{b2}(\rm 1P)$ (bottomonium); and (c) $B_c(\rm 1S)$, ~$B_c(\rm 2S)$,~ $B_{c}(\rm 1P)$ ($B_c$ meson). The equivalent UV cutoff for $N_{\max}=L_{\max}=8$ is $\mu_{c\bar c}\approx 2.8\,\mathrm{GeV}$, and for $N_{\max}=L_{\max}=32$ the UV cutoffs are  $\mu_{b\bar b}\approx 7.9\,\mathrm{GeV}$, $\mu_{b\bar c}\approx 6.77\,\mathrm{GeV}$, respectively.}
\label{fig_1}
\end{figure} 
 \begin{figure*}
 \subfloat[$N_{\max}=L_{\max}=8$  \label{fig:cj8}]{
 \includegraphics[width=0.44\textwidth]{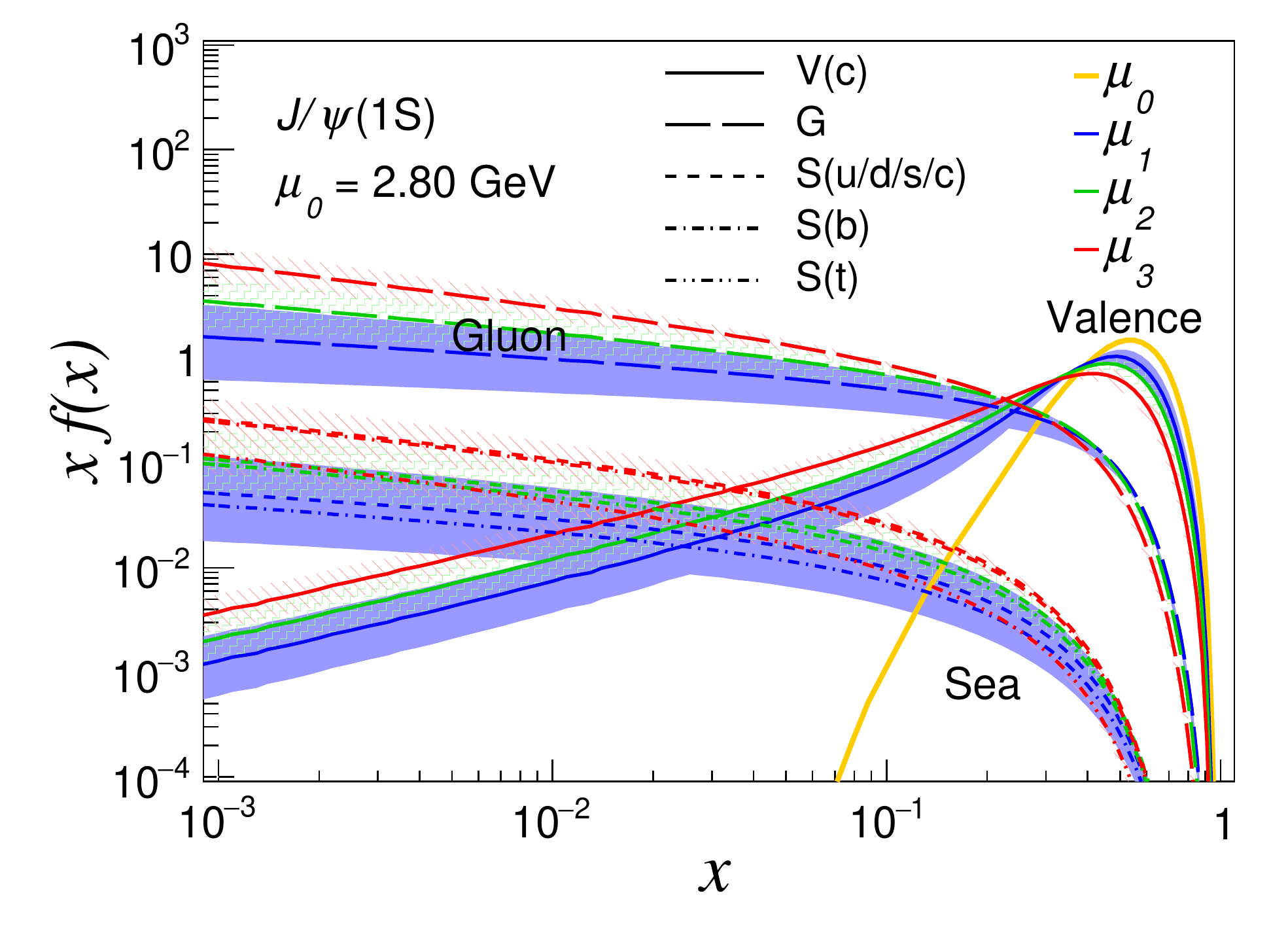}
}\quad
 \subfloat[$N_{\max}=L_{\max}=8$  \label{fig:cj8m}]{
 \includegraphics[width=0.44\textwidth]{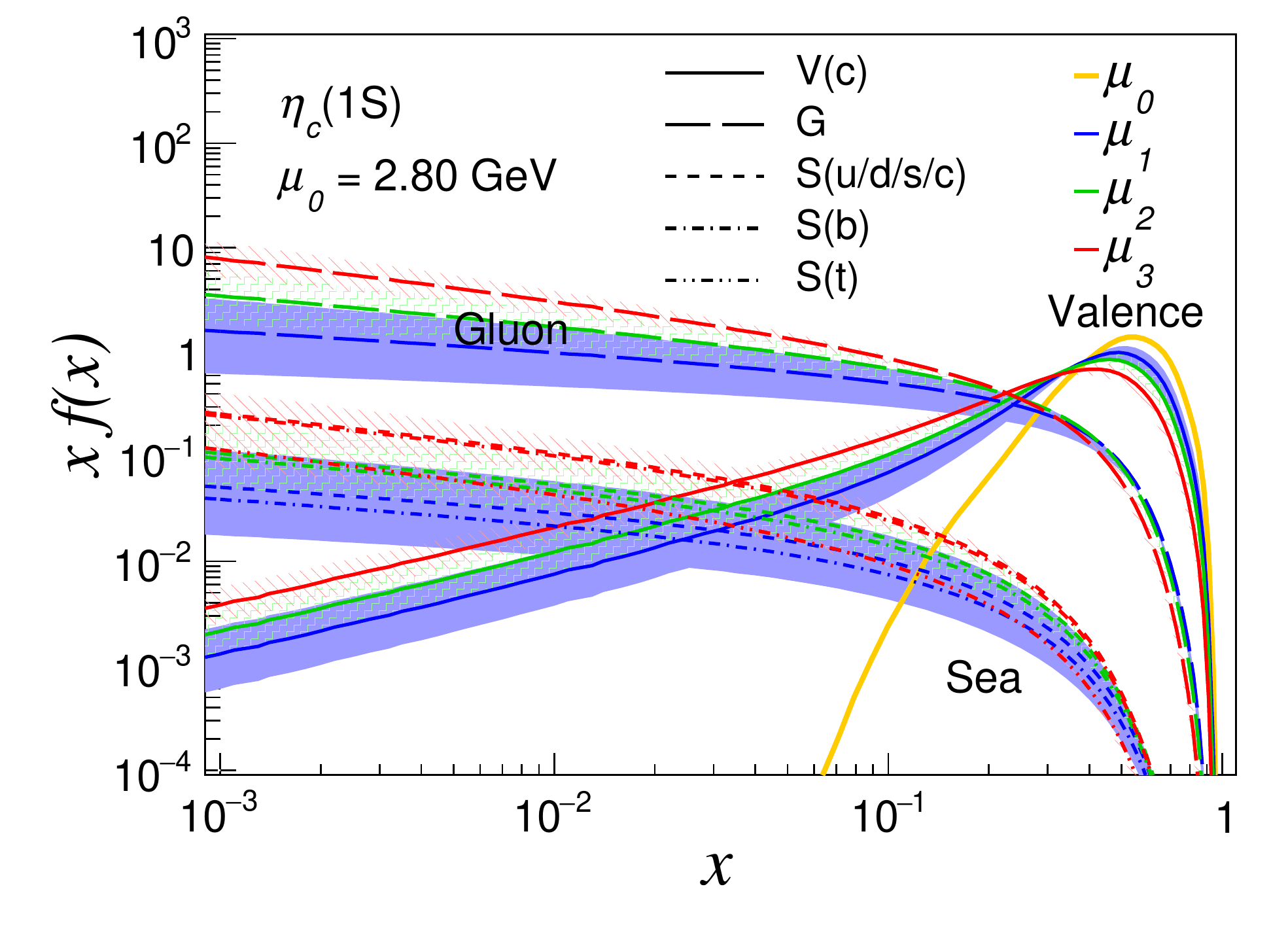}
}\\
\subfloat[$N_{\max}=L_{\max}=8$  \label{fig:cj8}]{
 \includegraphics[width=0.44\textwidth]{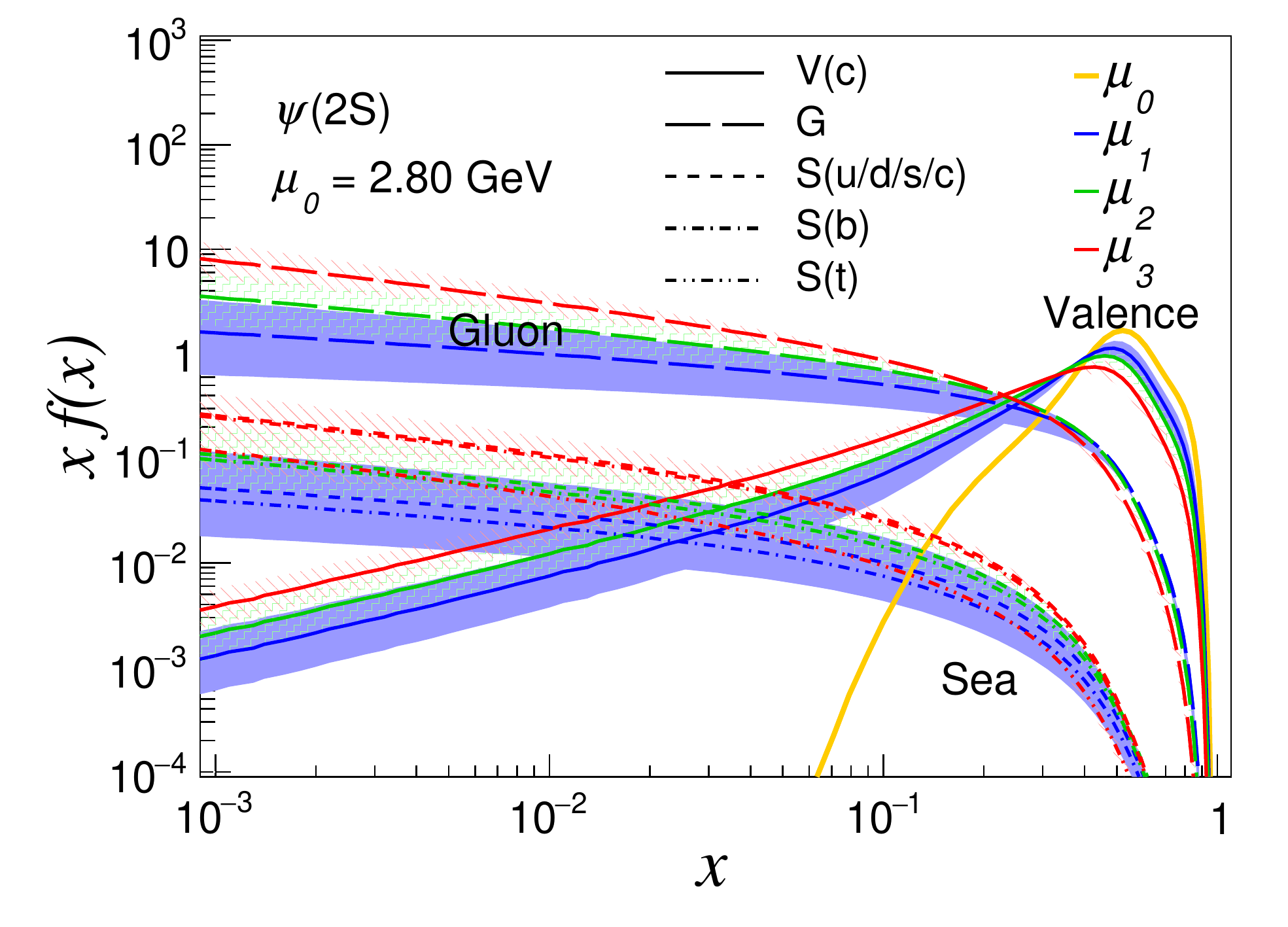}
}\quad
 \subfloat[$N_{\max}=L_{\max}=8$  \label{fig:cj8m}]{
 \includegraphics[width=0.44\textwidth]{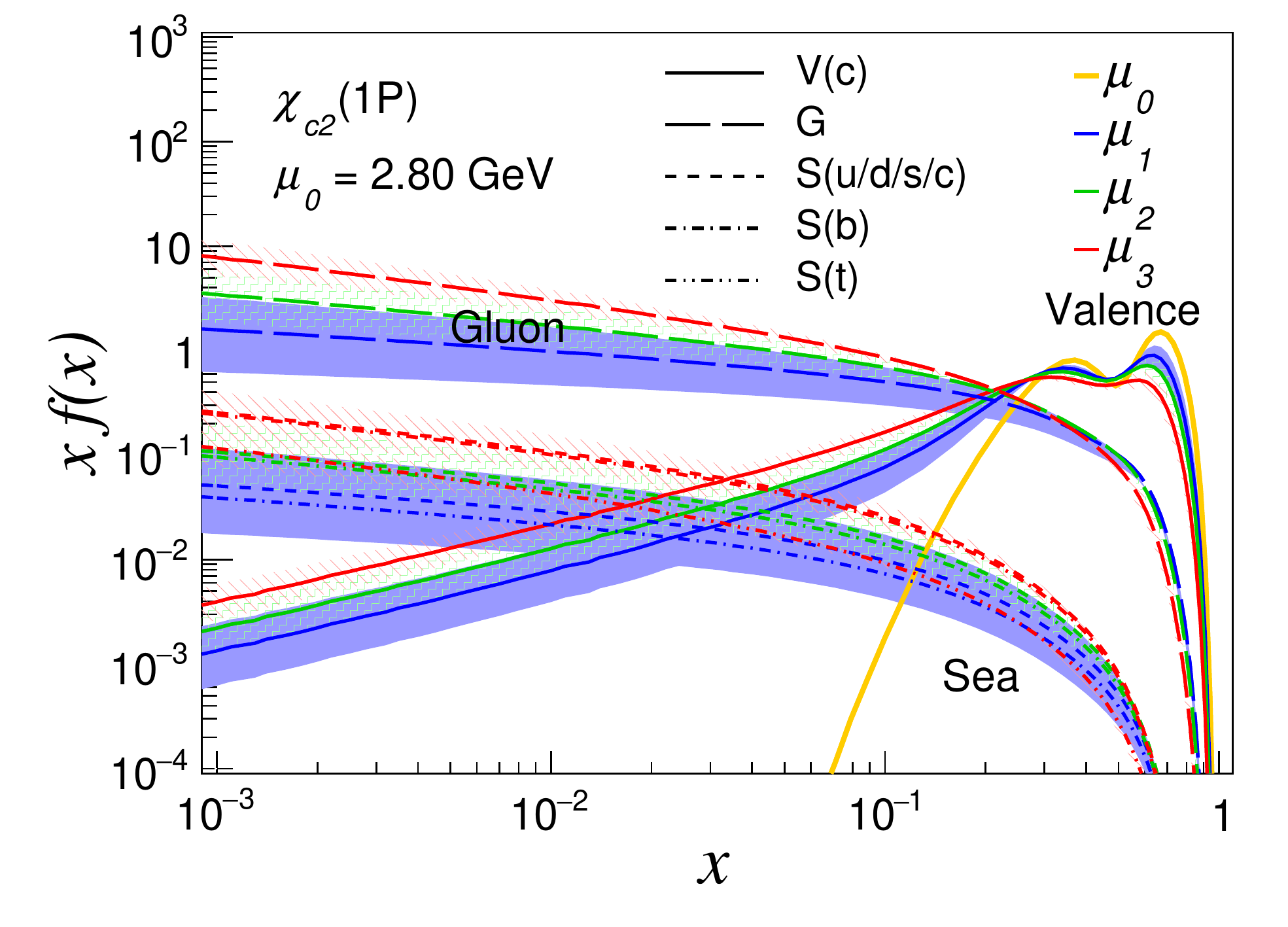}
}
\caption{The $x$-PDFs of charmonia: (a) $J/\psi(1\rm S)$, (b) $\eta_c(1\rm S)$, (c) $\psi(2\rm S)$, and (d) $\chi_{c2}(1\rm P)$ as a function of $x$ at different final scales $\mu$. The initial scale of the PDFs for the basis truncation $N_{\rm max}=8$  is the UV cutoff $\mu_0=2.80$ GeV. The bands represent the range of the distributions for the initial scales $\mu_0={\rm m}_q$ to $2\mu_h$. The lines with different color correspond to the different final scales: $\mu_1=20$ GeV (blue), $\mu_2=80$ GeV (green), and $\mu_3=1500$ GeV (red). The solid, thick long-dashed, dashed, dashed-dot, and dashed double-dot lines represent the $x$-PDFs of the valence quark, gluon, sea quark ($u/d/s/c$), sea quark ($b$), and sea quark ($t$), respectively.}
\label{fig_jpsi1s}
\end{figure*} 

\section{Parton distribution functions of heavy mesons}\label{pdf}
\subsection{The initial PDFs of heavy mesons}\label{evolution}
LFWFs play a central role in evaluating hadronic observables and light-cone distributions, and are an essential tool for investigating exclusive processes in DIS. PDFs control the inclusive processes at large momentum transfer~\cite{Lepage:1980fj}. The quark PDF, $q(x, \mu)$, is the probability of finding a collinear quark carrying momentum fraction $x$ up to scale $\mu$. In the light-front formalism, the PDF of the meson state with $m_J=0$ can be evaluated by integrating out the transverse momentum of the squared wave function within the two-body approximation \cite{Li:2017mlw}:
\begin{equation}
q(x, \mu) = \frac{1}{x(1-x)} \sum_{\lambda_{q}, \lambda_{\bar q}}\int\limits^{\mathclap{\lesssim\mu^2}} \frac{\dd^2\vec k_\perp}{2(2\pi)^3} \big| \psi_{\lambda_{q}\lambda_{\bar q}}(\vec k_\perp, x) \big|^2.\label{pdf_definition}
\end{equation}
The PDF and its first moment are normalized to unity and within the two-body approximation one can write
\begin{align}
\begin{split}
 \int_0^1 \dd x\, q(x, \mu) &= 1, \\
  \int_0^1  \dd x\, x\big [ q(x, \mu)+{\bar q}(x, \mu)\big] &= 1.
\end{split}
\end{align}
Note that each basis state is normalized to unity and satisfies each of these sum rules. Hence our eigenstates, which are normalized superpositions of these basis states, also satisfy these sum rules. These sum rules are therefore explicitly satisfied for all choices of basis space cutoffs.

Using the LFWFs mentioned as in Eq. (\ref{eqn:basis_representation}), the transverse integral in Eq. (\ref{pdf_definition}) can be carried out since,
\be
&&\int \frac{\dd^2\vec k_\perp}{(2\pi)^2}\phi_{nm}\left(\vec k_\perp/\sqrt{x(1-x)}\right)\phi^*_{n'm'}\left(\vec k_\perp/\sqrt{x(1-x)}\right)\nonumber\\
&&= x(1-x)\delta_{n,n'}\delta_{m,m'}.
\ee
The PDFs in the basis function representation then follow as:
\be
q(x,\mu)=&&\frac{1}{4\pi} \sum_{\lambda_{q},\lambda_{\bar q}}\sum_{n,m,l,l'} \langle n, m, l, \lambda_q, \lambda_{\bar q} | \psi^J_{m_J}\rangle\nonumber\\
&&\times \langle \psi^J_{m_J} |  n, m, l', \lambda_q, \lambda_{\bar q}\rangle \chi_l(x) \chi_{l'}(x).
\ee
Here, we consider eight heavy quarkonium states which include two scalar particles ($\eta_c(1\rm S),~\eta_b(1\rm S)$), four vector particles, and two tensor particles with $m_J=0$. $J/\psi(1\rm S)$, $\psi(2\rm S)$ and $\Upsilon(1\rm S)$, $\Upsilon(2\rm S)$ are the radially excited states with $J=1$, whereas $\chi_{c2}(1\rm P)$ and $\chi_{b2}(1\rm P)$ are the tensor states for $J=2$. In order to study the PDFs of heavy mesons with unequal quark masses,  we also consider three $B_c$ meson states: $B_c(\rm 1S)$, $B_c(\rm 2S)$ and $B_c(\rm 1P)$.   
It should be noted that for the systems with  $J \geq1$, there exists more than one PDF depending on the polarizations of the active quark as well as the system. For example, for spin-1 hadrons, there is also the well known tensor polarized PDF which is defined as $b_1(x)=(2q_{m_J=0}(x)-q_{m_J=1}(x)-q_{m_J=-1}(x))/4$.
The tensor polarized PDF $b_1(x)$, being sensitive to the parton's orbital angular momentum, attracts significant theoretical as well as experimental attention~\cite{Hoodbhoy:1988am,Close:1990zw,Umnikov:1996qv,Detmold:2005iz,Kumano:2010vz,Islam:2012zua,Miller:2013hla,Kumano:2016ude,Cosyn:2017fbo,Ninomiya:2017ggn,Airapetian:2005cb}. We focus on the unpolarized PDFs in this work. A detailed analysis of all other PDFs of vector and tensor particles in our BLFQ approach will be reported in a future study.

In Fig. \ref{fig:inicc8} and Fig. \ref{fig:inicc32}, we show the unpolarized PDFs of charmonium ($\eta_c(1\rm S)$, $J/\psi(1\rm S)$, $\psi(2\rm S)$, $\chi_{c2}(1\rm P)$) and bottomonium ($\eta_b(1\rm S)$, $\Upsilon(1\rm S)$, $\Upsilon(2\rm S)$, $\chi_{b2}(1\rm P)$) states evaluated using the LFWFs for $N_{\max}=L_{\max}=8$ and $N_{\max}=L_{\max}=32$, respectively. The chosen basis functions: $N_{\max}=8$ for charmonium and $N_{\max}=32$ for bottomonium roughly correspond to UV regulator $\mu_h=\Lambda_{UV}\sim\kappa\sqrt{N_{\rm max}}\approx 1.7 {\rm m}_q$ for quarkonia and ${\rm m}_h+{\rm m}_{\bar{h}}$ for $B_c$. Our choice of the regulators of the basis functions is motivated by the competition between the necessities for both a better basis resolution and a lower UV scale since the present model does not incorporate radiative corrections \cite{Li:2017mlw}. It is interesting to note that the PDFs for $\eta_c(1\rm S)$ and $J/\psi(1\rm S)$ exhibit a similar behavior; however, the PDFs for $\psi(2\rm S)$ and $\chi_{c2}(1\rm P)$  show a distinctly different behavior from the other two. There appear to be ripples on the downward slopes of the PDFs for  $\psi(2\rm S)$ whereas there is a dip at $x=1/2$ in the PDF of $\chi_{c2}(1\rm P)$ which may be expected from contributions of longitudinally excited basis functions. The qualitative behavior of bottomonium $1\rm S~ (\eta_b~\rm{and}~\Upsilon)$, $2\rm S~ (\Upsilon)$ and $1\rm P~ (\chi_{b2})$ states is more or less the same as charmonium $1\rm S~ (\eta_c~\rm{and}~J/\psi)$, $2\rm S~(\psi)$, and $1\rm P~(\chi_{c2})$ states. However, the width of the distributions for charmonium is larger compared to that for bottomonium. This is expected due to the smaller masses of charmonium states than the bottomonium masses. Furthermore, at a larger mass scale, the running coupling is smaller which leads to the smaller kinetic energy in bottomonium. Thus, the probability of carrying small longitudinal momentum by the quark/antiquark in bottomonium is always small and the probability is higher when they share equally momentum. Effectively, the momentum space distributions are narrower in bottomonium systems than that in charmonium. 

The valence quark PDFs of three $B_c$ meson states at the chosen hadronic scale are shown in Fig. \ref{fig:inibb8}. The peaks of the charm quark PDFs appear at lower $x$, whereas due to the heavier mass, the bottom quark distributions have the peaks at higher $x$. We also observe that although the PDFs of $B_c(\rm 1S)$ and $B_c(\rm 2S)$ show a similar behavior, the PDF of $B_c(\rm 1P)$ exhibits a somewhat different character reminiscent of the $1 {\rm P}$ states shown in panels (a) and (b) of Fig. \ref{fig_1}. Note that $B_c(\rm 2S)$ is broader than $B_c(\rm 1S)$ as may be expected. This is to be compared with charmonium state $\psi(2\rm S)$ and bottomonium state $\Upsilon(2\rm S)$ which, in addition, have ripples on the downward slopes.  


\subsection{QCD evolution of heavy meson PDFs}\label{evolution}
By performing the QCD evolution, the valence quark PDFs at high  scale can be obtained with the input valence PDFs at the initial scale.  The  DGLAP \cite{Dokshitzer:1977sg,Gribov:1972ri,Altarelli:1977zs} equation, which bridges PDFs between a final scale and an initial scale, is given by, 
\begin{equation}
\begin{aligned}
&\frac{\partial}{\partial ~\rm{ln} ~\mu^2}\left( \begin{array}{c}
q(x,\mu)\\g(x,\mu)\end{array}\right)\\
=&\frac{\alpha_s(\mu^2)}{2\pi}\int_x^1\frac{dy}{y}\left( \begin{array}{cc}P_{qq}(x/y)&P_{qg}(x/y)\\
P_{gq}(x/y)&P_{gg}(x/y)\end{array}\right)\left( \begin{array}{c}
q(y,\mu)\\g(y,\mu)\end{array}\right),
\end{aligned}
\end{equation}
where $P_{qq}(z)$, $P_{qg}(z)$, $P_{gq}(z)$ and $P_{gg}(z)$ are the splitting kernels. Here, we adopt the DGLAP equations of QCD up to NNLO, to evolve our PDFs from the model scales ($\mu_{0}\gg \Lambda_{\rm {QCD}}$), to higher scales ($\mu$). The QCD evolution allows quarks to emit and absorb gluons, with the emitted gluons allowed to create quark-antiquark pairs as well as additional gluons. In this picture, the sea quark and gluon components of the constituent quarks are revealed at higher scale through QCD evolution. Here, we use the higher order perturbative parton evolution toolkit (HOPPET) to numerically solve the NNLO DGLAP equations~\cite{Salam:2008qg}. The large mass scales of heavy mesons provide grounds for the usage of perturbative evolution.

 \begin{figure*}

 \subfloat[$N_{\max}=L_{\max}=32$  \label{fig:bu32}]{
 \includegraphics[width=0.44\textwidth]{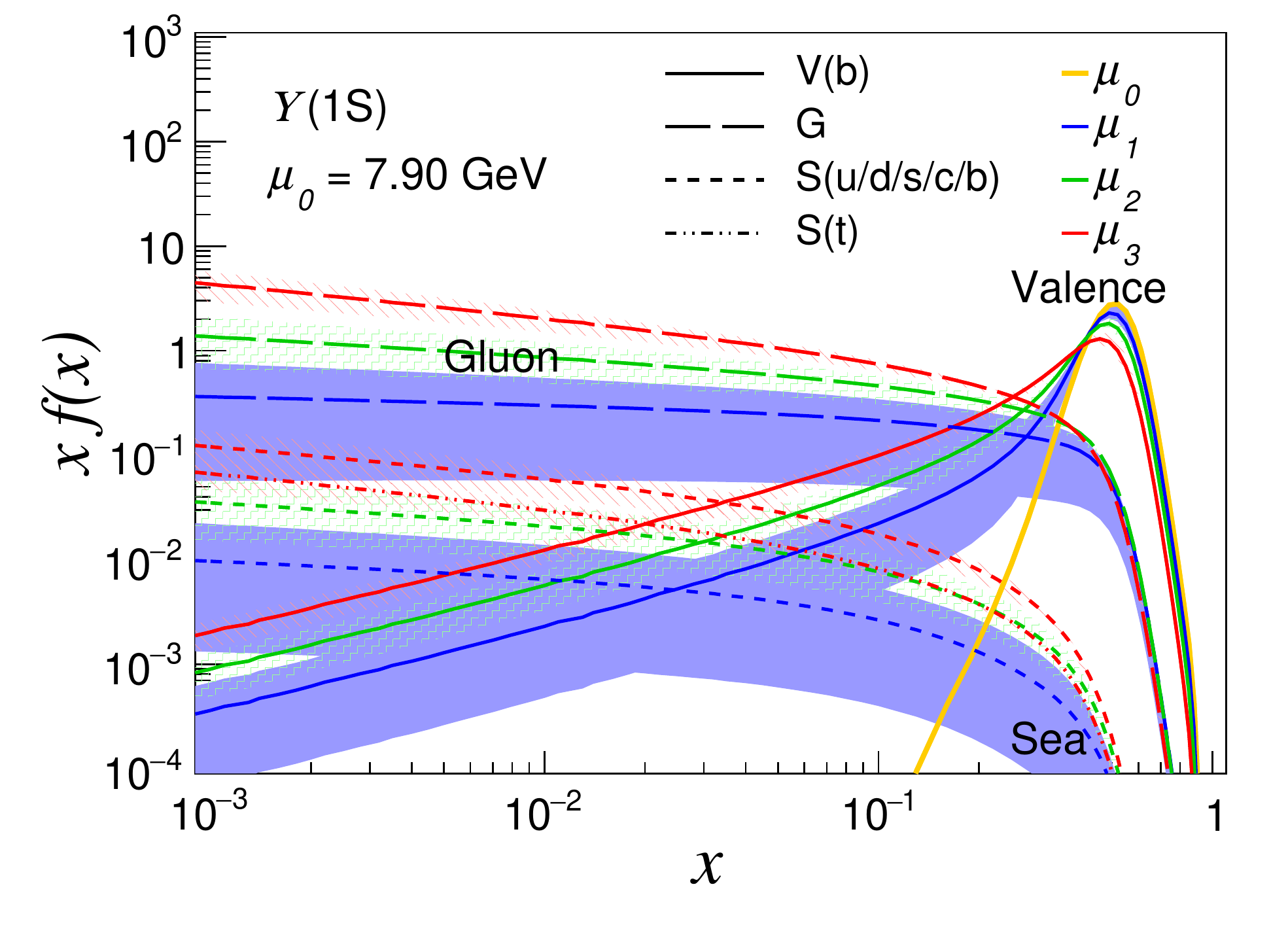}
}\quad 
\subfloat[$N_{\max}=L_{\max}=32$  \label{fig:bu32m}]{
 \includegraphics[width=0.44\textwidth]{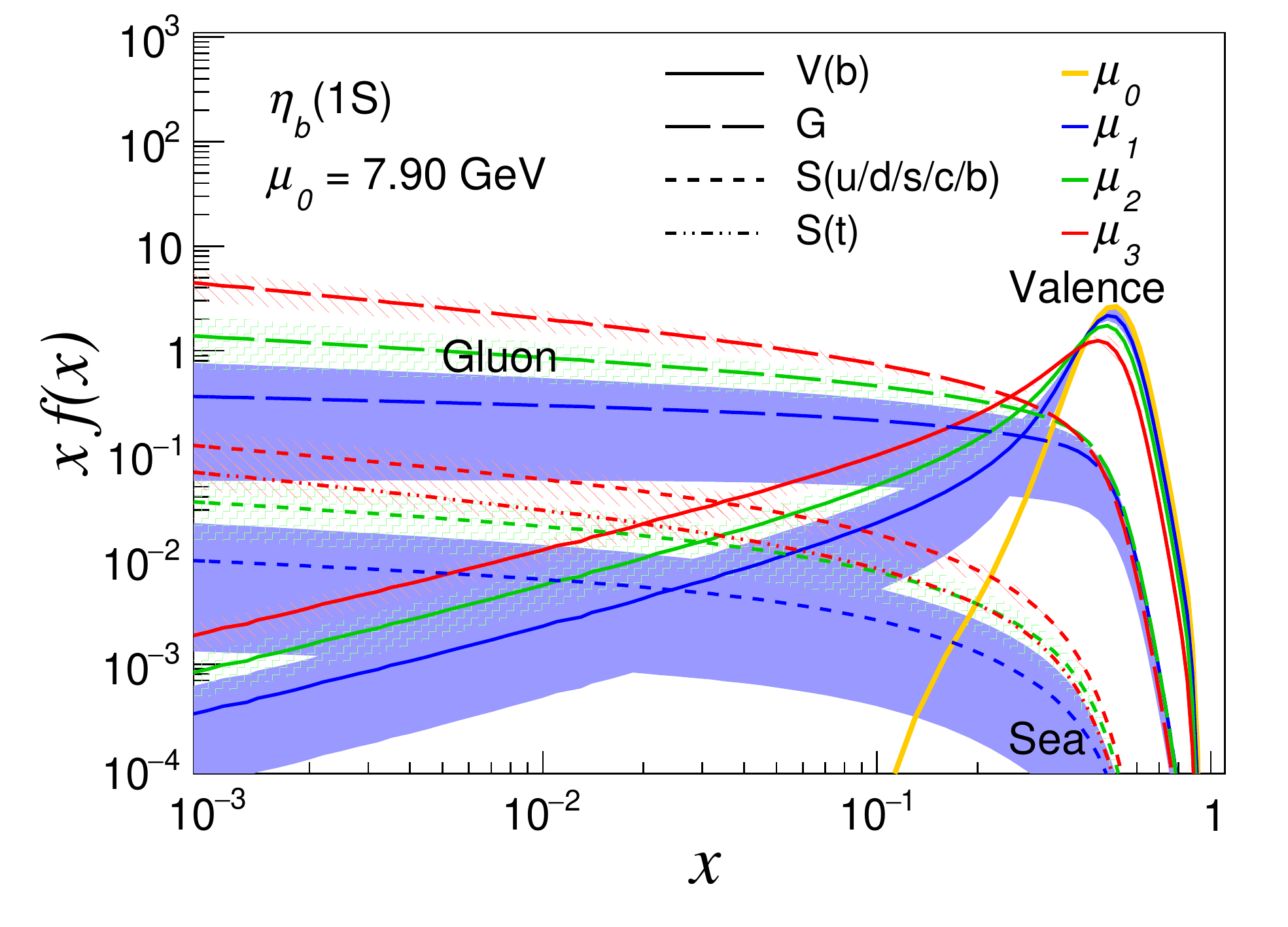}
}\\
 \subfloat[$N_{\max}=L_{\max}=32$  \label{fig:bu32}]{
 \includegraphics[width=0.44\textwidth]{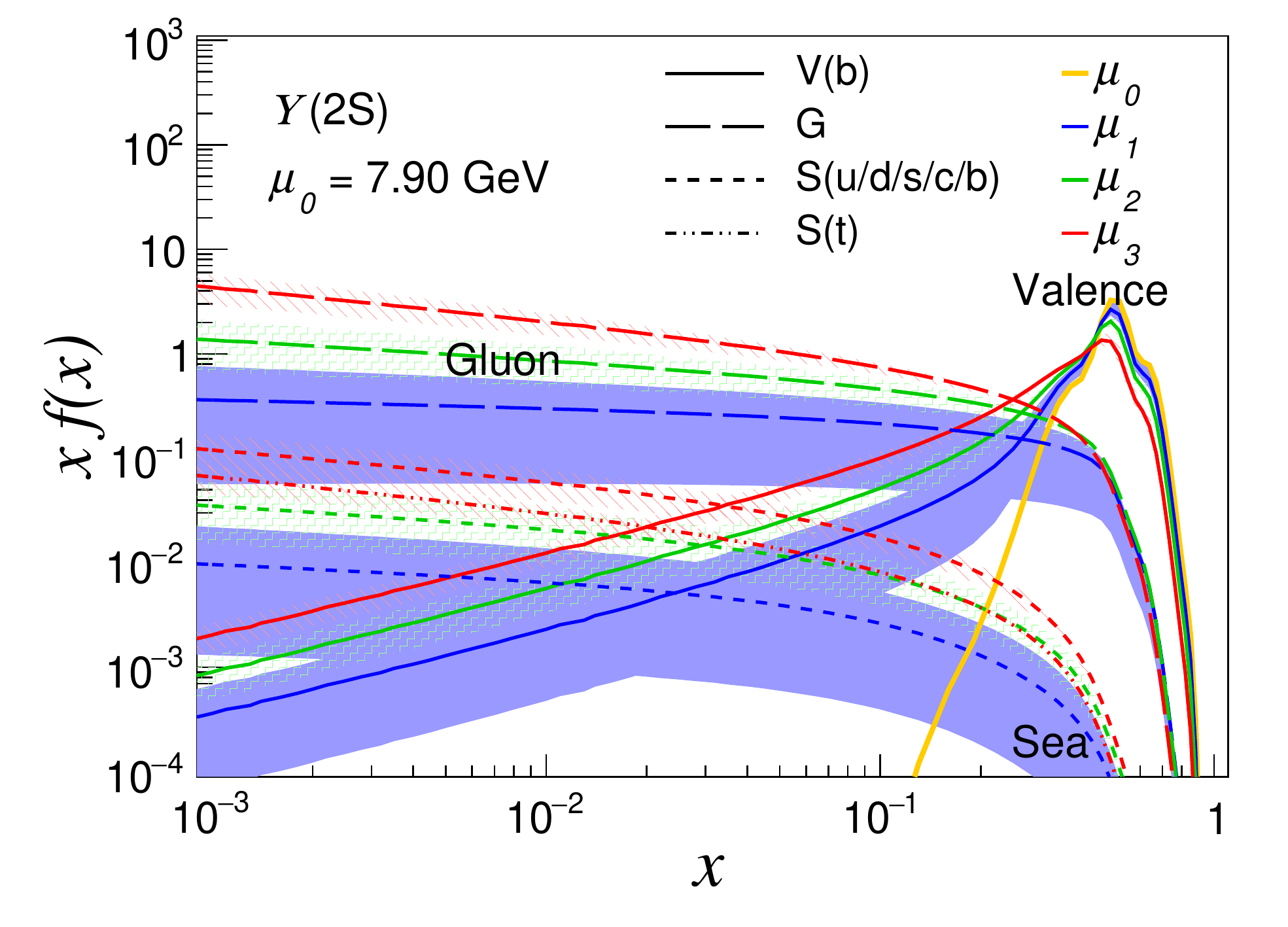}
}\quad 
\subfloat[$N_{\max}=L_{\max}=32$  \label{fig:bu32m}]{
 \includegraphics[width=0.44\textwidth]{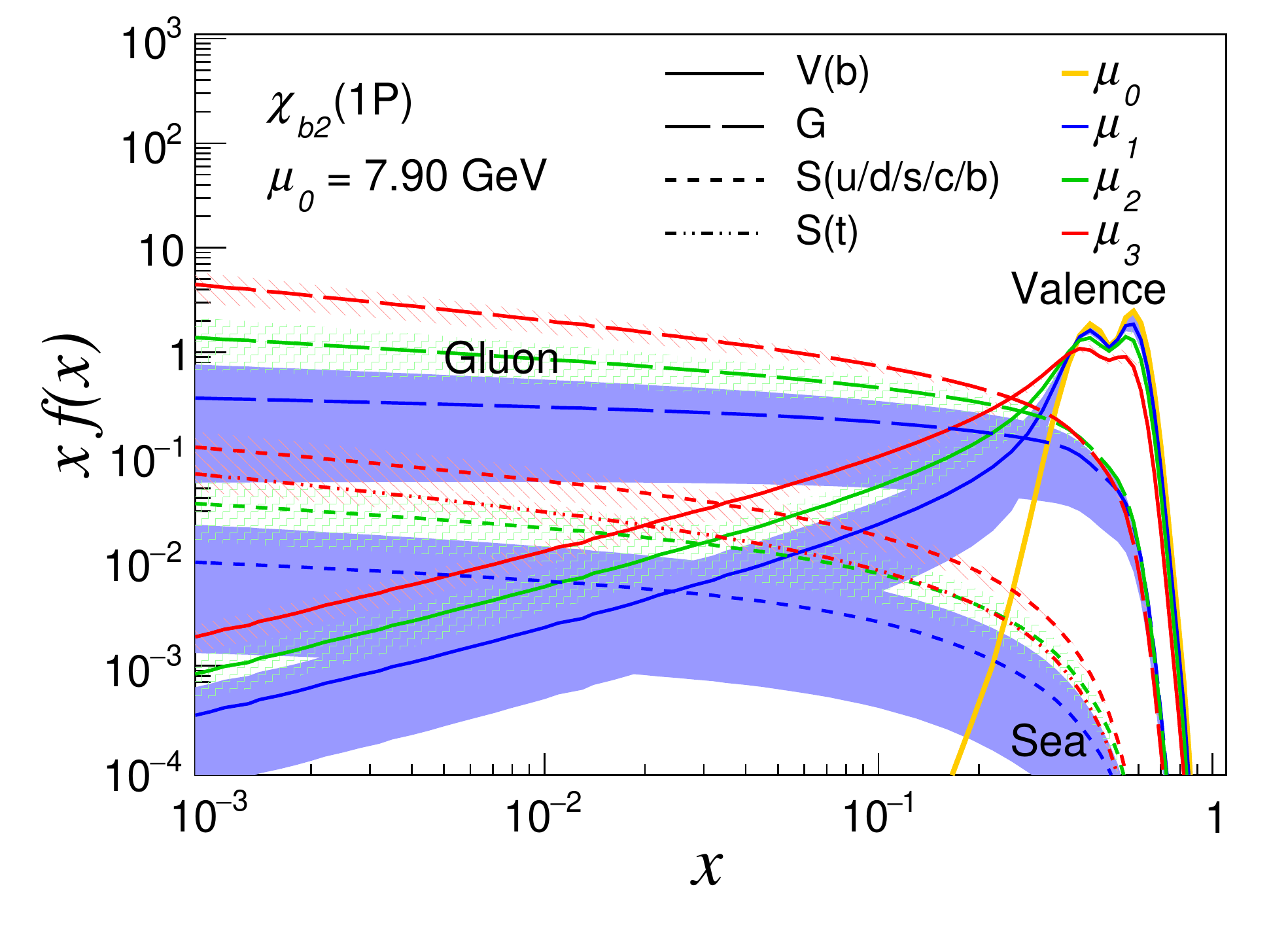}
}
\caption{The plots of (a) the $x$-PDFs of the bottomonia: (a) $\Upsilon(1\rm S)$, (b) $\eta_b(1\rm S)$, (c) $\Upsilon(2\rm S)$, and (d) $\chi_{b2}(1\rm P)$ as a function of $x$ at different final scales $\mu$. The initial scale of the PDFs for the basis truncations $N_{\rm max}=32$ is the UV cutoff $\mu_0=7.90$ GeV.  The bands represent the range of the distributions for the initial scales $\mu_0={\rm m}_q$ to $2\mu_h$. The lines with different color correspond to the different final scales: $\mu_1=20$ GeV (blue), $\mu_2=150$ GeV (green), and $\mu_3=1500$ GeV (red).
 The solid, thick long-dashed, dashed, and dashed double-dot lines represent to the $x$-PDFs of valence quark, gluon, sea quark ($u/d/s/c/b$), and sea quark ($t$), respectively.}
\label{fig_upsilon1sm}
\end{figure*}

We evolve the PDFs of charmonia: $\eta_c(1\rm S)$, $J/\psi(1\rm S)$, $\psi(2\rm S)$, $\chi_{c2}(1\rm P)$, bottomonia: $\eta_b(1\rm S)$, $\Upsilon(1\rm S)$, $\Upsilon(2\rm S)$, $\chi_{b2}(1\rm P)$, and $B_c$ mesons: $B_c (\rm 1S)$, $B_c(\rm 2S)$, $B_c(\rm 1P)$ obtained in the basis function representation. As mentioned above, the initial scale $\mu_0$ of the PDF is chosen as a low UV cutoff $\mu_h$ to suppress the radiative corrections. Specifically, we adopt the UV cutoffs for  $N_{\rm max}=8$ for charmonia and  $N_{\rm max}=32$ for bottomonia as well as $B_c$ mesons \cite{Li:2017mlw, Tang:2018myz}. For a comprehensive study, we also vary the initial scales by choosing $\mu_0={\rm m}_q$ and $\mu_0=2\mu_h$. The difference in results is an indicator of the sensitivity with respect to the choice of the initial scale.
The initial scales are given in Table \ref{T1}. The PDFs are evolved to final scales $20$, $80$, and $1500$ GeV which are the relevant scales for the proposed Electron Ion Collider in China (EicC) \cite{Chen:2018wyz}, the electron-Relativistic Heavy Ion Collider (eRHIC) \cite{Aschenauer:2014cki}, and the proposed Large Hadron Electron Collider (LHeC) \cite{AbelleiraFernandez:2012cc}, respectively. Here, we consider the range $x\ge 10^{-3}$. We expect that at low initial scale the DGLAP evolution with a leading twist is not sufficient at low $x$~\cite{Zhu:2010qa,Boroun:2010zza} and one needs to consider the higher twist corrections in the DGLAP equation~\cite{Boroun:2013mgv,Boroun:2010zz,Devee:2014fna,Phukan:2017lzp,Rezaei:2010zz,Boroun:2009zzb,Lalung:2017omk}. It should be mentioned here that there is also uncertainty from the longitudinal basis resolution within the initial scale PDFs. The uncertainty is proportional to $1/L_{\rm max}$ and would propagate to the PDFs at final scales.

 \begin{figure*}
 \subfloat[$N_{\max}=L_{\max}=32$  \label{fig:bc1sa}]{
 \includegraphics[width=0.44\textwidth]{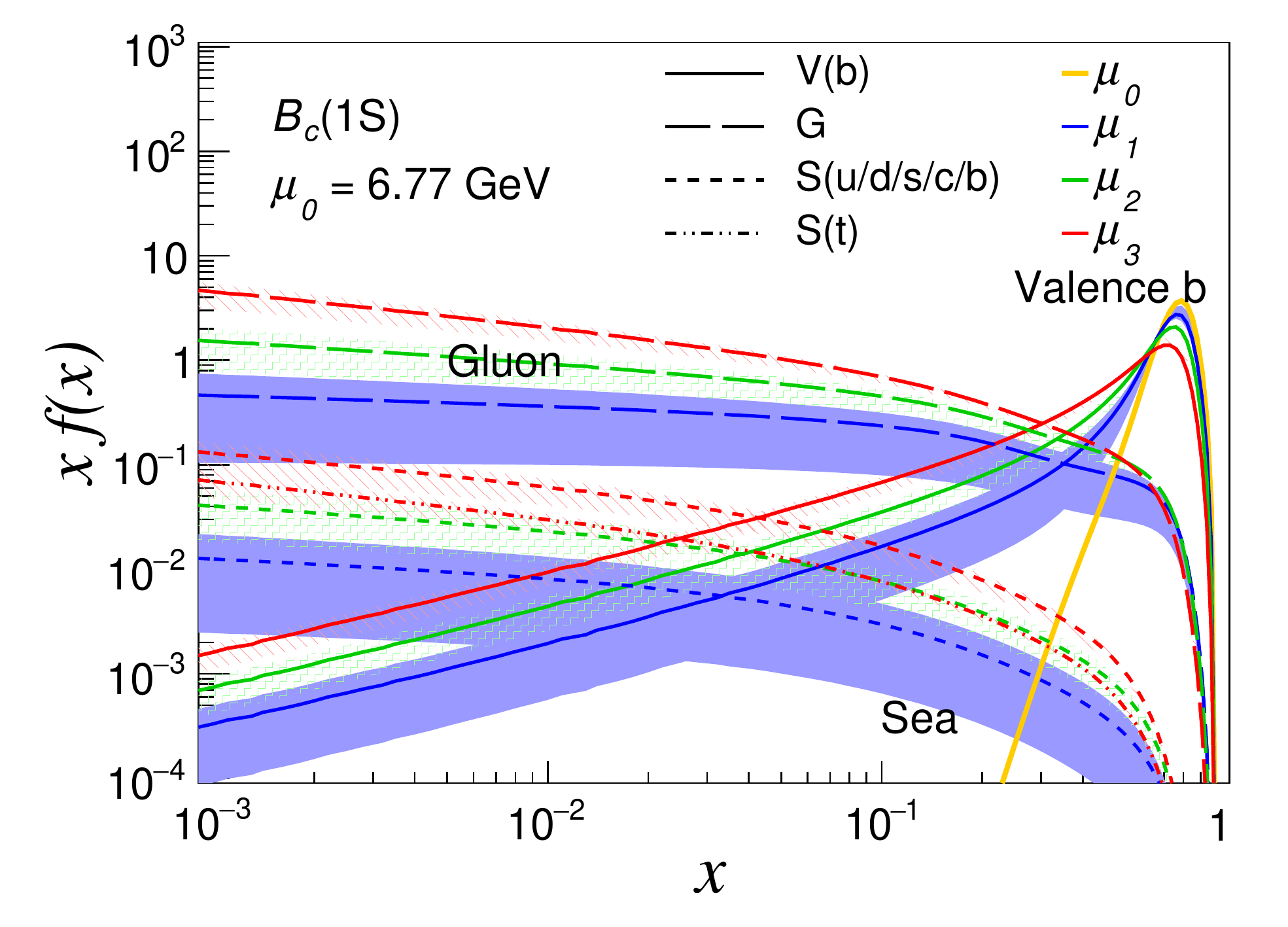}
}\quad
 \subfloat[$N_{\max}=L_{\max}=32$  \label{fig:bc1sc}]{
 \includegraphics[width=0.44\textwidth]{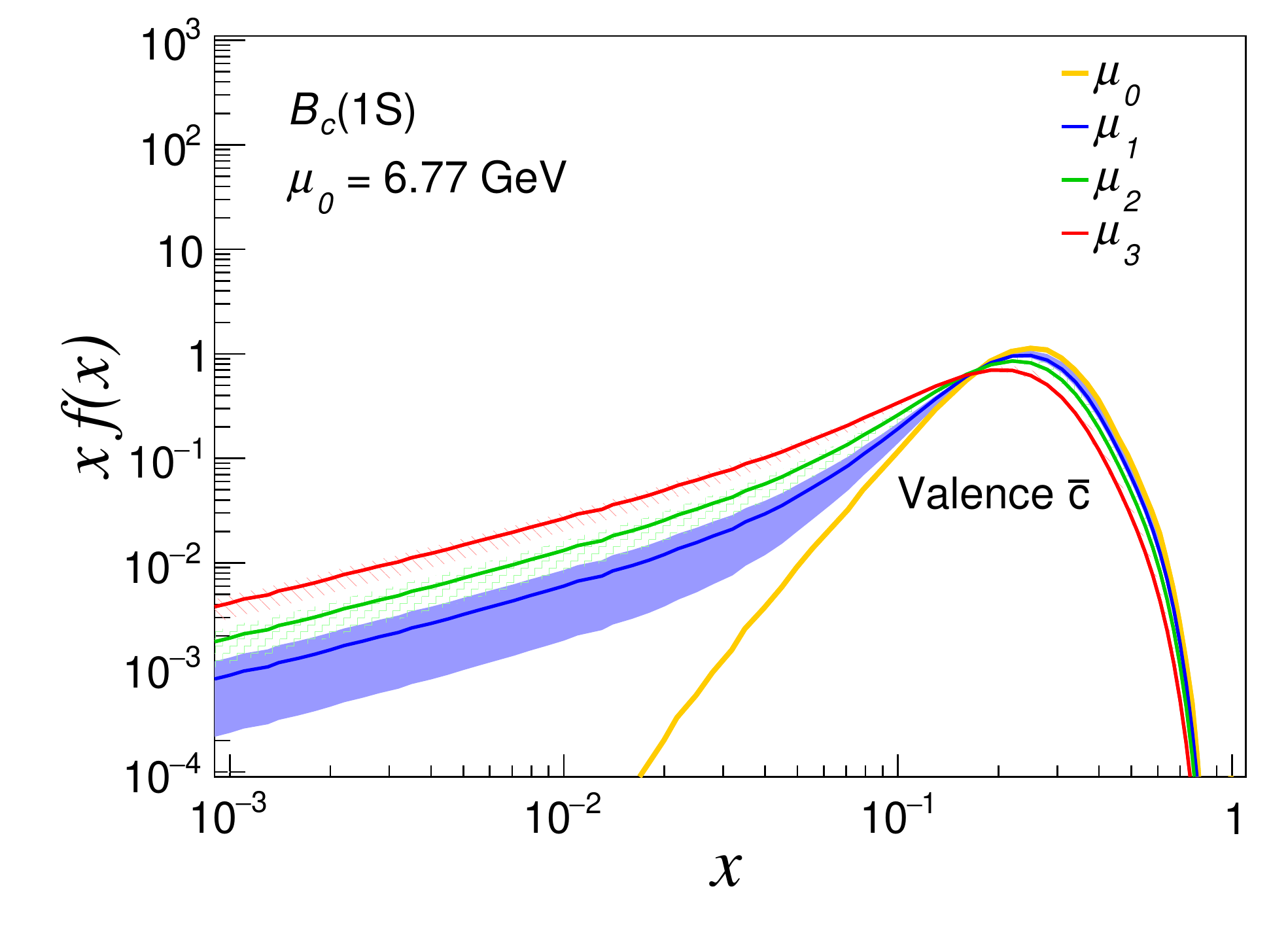}
} 

\caption{(a) and (b) The $x$-PDFs of the $B_c(1\rm S)$ as a function of $x$ at different final scales $\mu$ (a) valence $b$ quark, (b) valence $\bar{c}$ quark. The initial scale is $\mu_0=6.77$ GeV and the truncation parameter is $N_{\rm max}=32$. The bands represent the range of the distributions for the initial scales $\mu_0={\rm m}_q$ to $2\mu_h$. The lines in (a) and (b) correspond to the same parameter values as presented in the caption to Fig. \ref{fig_upsilon1sm}.} 
\label{fig_bc1s}
\end{figure*} 

 \begin{figure*}
 \subfloat[charmonia \label{fig:bc2sa}]{
 \includegraphics[width=0.44\textwidth]{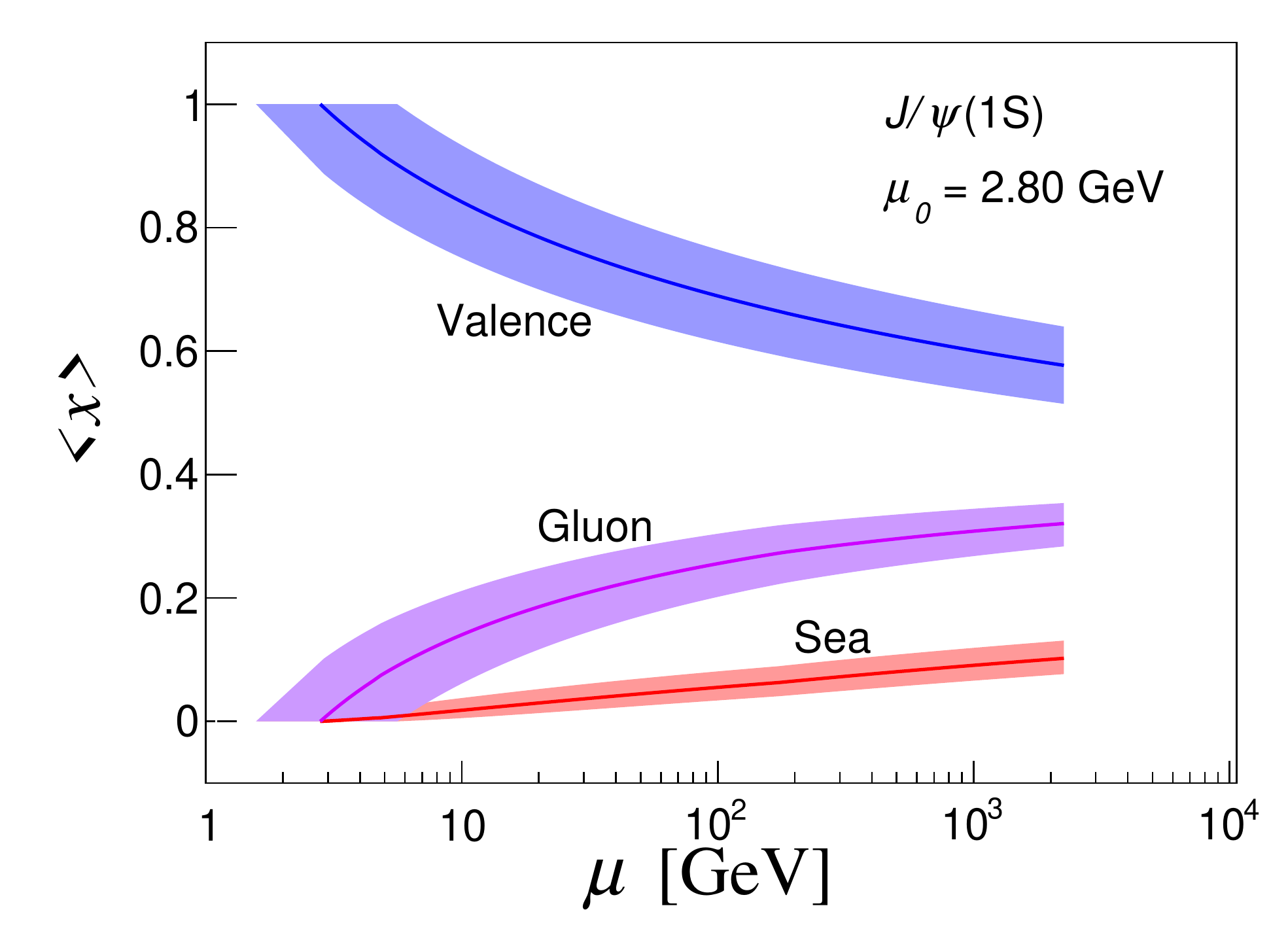}
}\quad
 \subfloat[bottomonia  \label{fig:bc2sc}]{
 \includegraphics[width=0.44\textwidth]{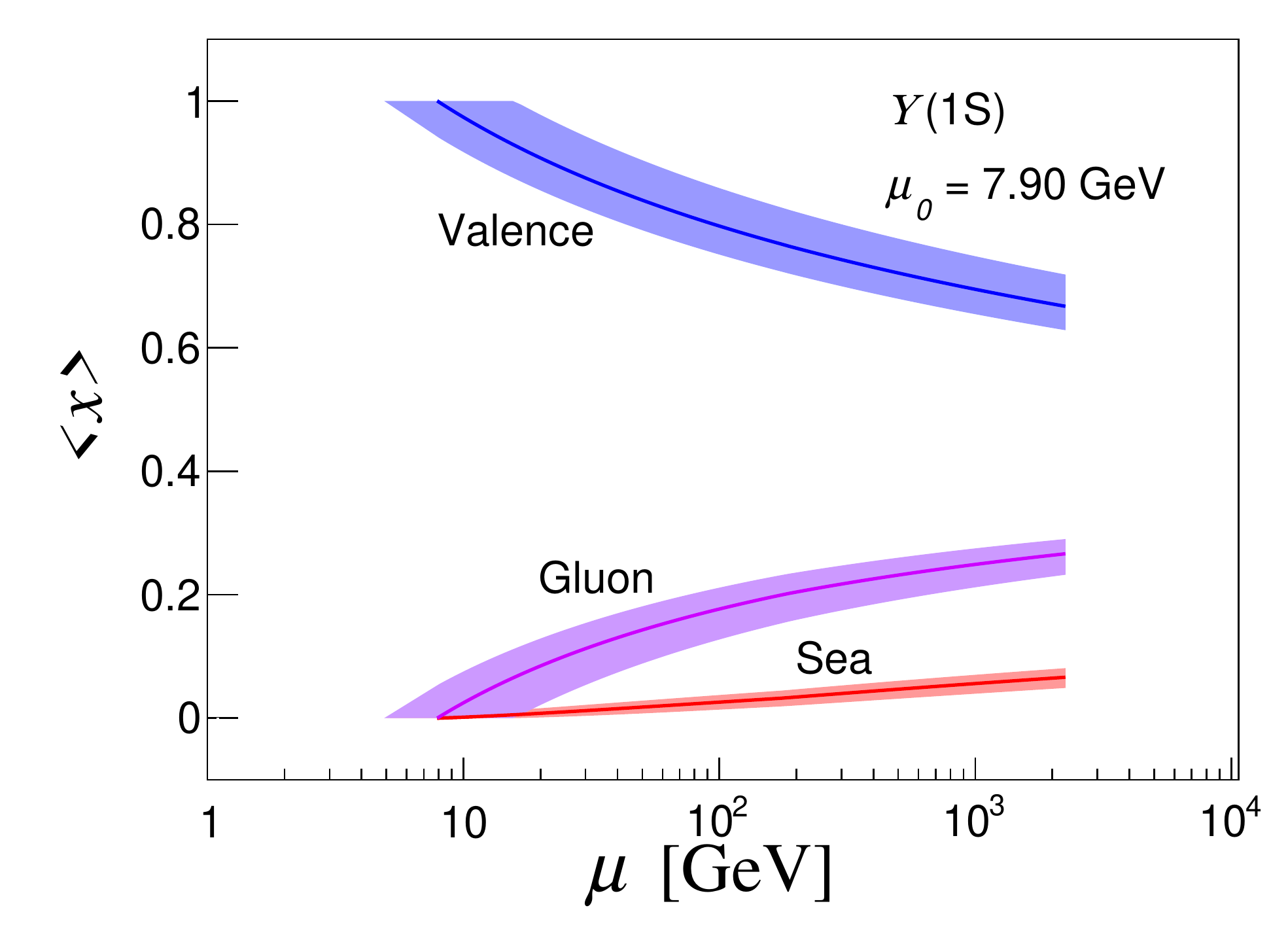}
} 

 \subfloat[$B_c$ mesons  \label{fig:bc2sm}]{
 \includegraphics[width=0.44\textwidth]{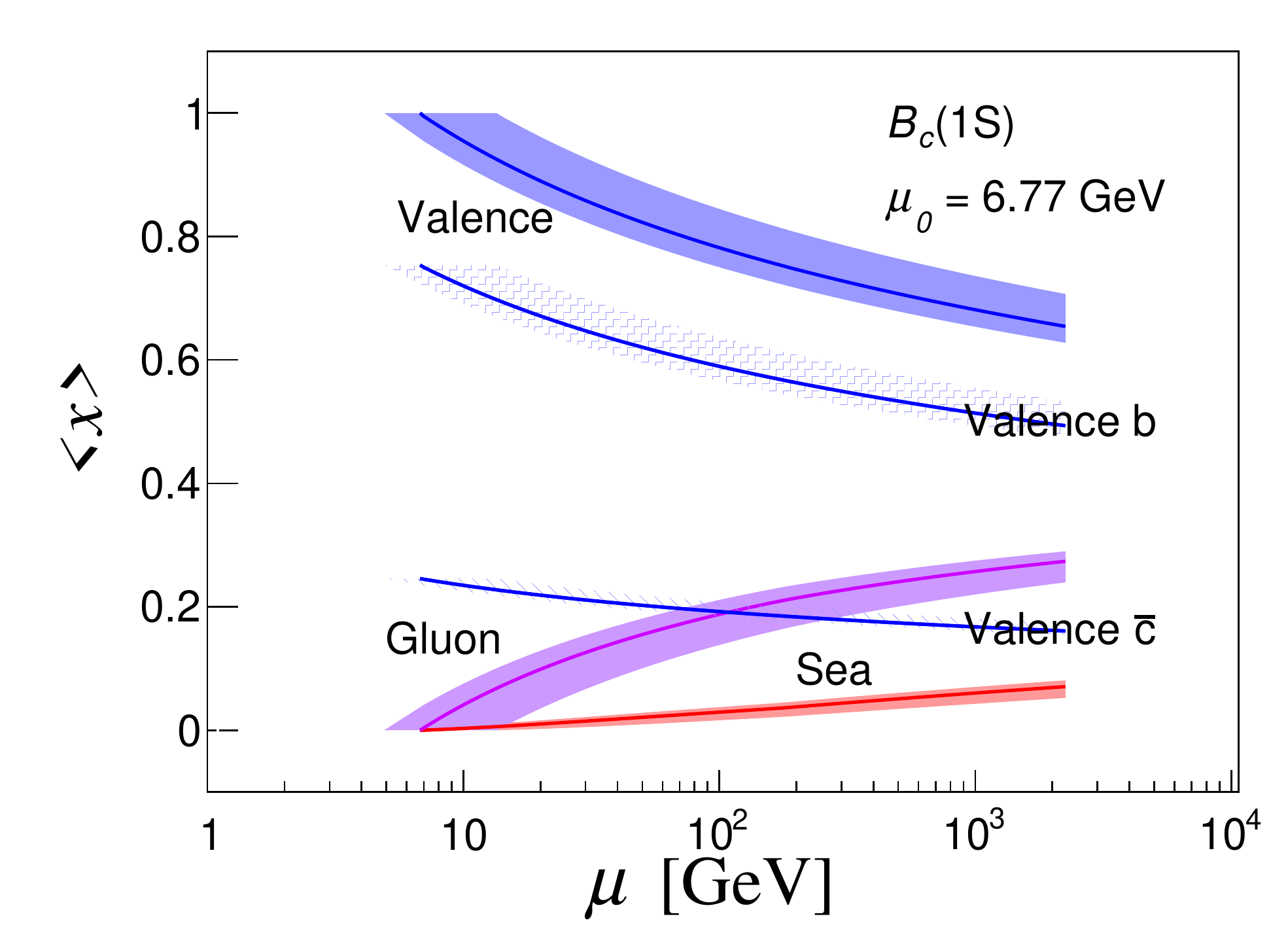}
}
\caption{The first moment of the PDFs of (a) charmonium: $J/\psi(1\rm S)$, (b) bottomonium: $\Upsilon(1\rm S)$, and (c) $B_c$ meson: $B_c(1\rm S)$ as function of the scale $\mu$. The bands represent the range of the distributions for the initial scales $\mu_0={\rm m}_q$ to $2\mu_h$. The lines with blue, purple, and red represent the first moments of valence quark, gluon, and sea quarks, respectively.}
\label{1st_moment}
\end{figure*} 

We demonstrate the evolution of the PDFs of the charmonia and bottomonia states from the initial scales to the relevant Electron-Ion Collider (EIC) scales 
in Fig. \ref{fig_jpsi1s} and Fig. \ref{fig_upsilon1sm}, respectively. 
We observe that for both the charmonium and the bottomonium, their valence quark distributions increase slowly at lower $x$ but decrease at higher $x$, e.g. $x>0.3$, with the scale evolution. The gluon and the sea quark PDFs at low $x$ increase much faster than the valence quark PDFs. Effectively, in the low $x$ region the distributions are mainly dominated by the gluon PDFs, whereas at large $x$ the valence quark dominates the distributions. 
 The PDFs for gluon, sea quark ($u/d/s/c$), sea quark ($b$), and sea quark ($t$) shown in these plots, are generated by the QCD evolution of the valence quark PDFs. The bands in those figures represent the range of the distributions for the initial scales $\mu_0={\rm m}_q$ to $2\mu_h$, while the lines correspond to the UV cutoff for the basis truncations chosen as the initial scales. The different scales of the PDFs have been represented by different colors in those figures (left panels). One can notice that the qualitative behavior of the gluon and the sea quarks PDFs obtained by the evolution in both the charmonia and the bottomonia states is very similar.

The evolution of the valence $b$ and $\bar{c}$ quark PDFs in the $B_c$ meson state, $B_c(\rm 1S)$ is demonstrated in Fig.~\ref{fig:bc1sa} and Fig.~\ref{fig:bc1sc}, respectively. The PDFs for gluon, sea quarks, generated by the QCD evolution of the valence quark PDFs are also presented in Fig.~\ref{fig:bc1sa}. 
Since the masses of the $b$ and $c$ quarks are very different, the peaks of their distributions appear at different $x$. 
Again, we observe that the gluon distribution dominates at low $x$ while, at large $x$, the distribution is dominated by the valence quark distribution. 
We notice that other two $B_c$ meson states, $B_c(\rm 2S)$ and $B_c(\rm 1P)$ exhibit a similar behavior as the PDF of $B_c(\rm 1S)$ in Fig.~\ref{fig_bc1s} with QCD evolution. 

The first moments of the corresponding PDFs of charmonium, bottomonium, and $B_c$ meson as functions of $\mu$ are shown in Fig.~\ref{1st_moment}. For the demonstration purposes, we consider the $J/\psi(1\rm S)$, $\Upsilon (1\rm S)$, and $B_c(1\rm S)$ states. We find that with increasing scale $\mu$, the momentum carried by the valence quark decreases and the contributions of the sea quarks and gluon to the total momentum increase. Due to heavier mass, at a particular scale, the momentum carried by $b$ quark is larger compared to $\bar{c}$ quark in the $B_c$ meson. However, the qualitative behaviors of the total moments of valence quarks, sea quarks, and gluons  in all heavy meson are alike.

\begin{figure*}[htp]
\begin{center}
\includegraphics[width=0.45\textwidth]{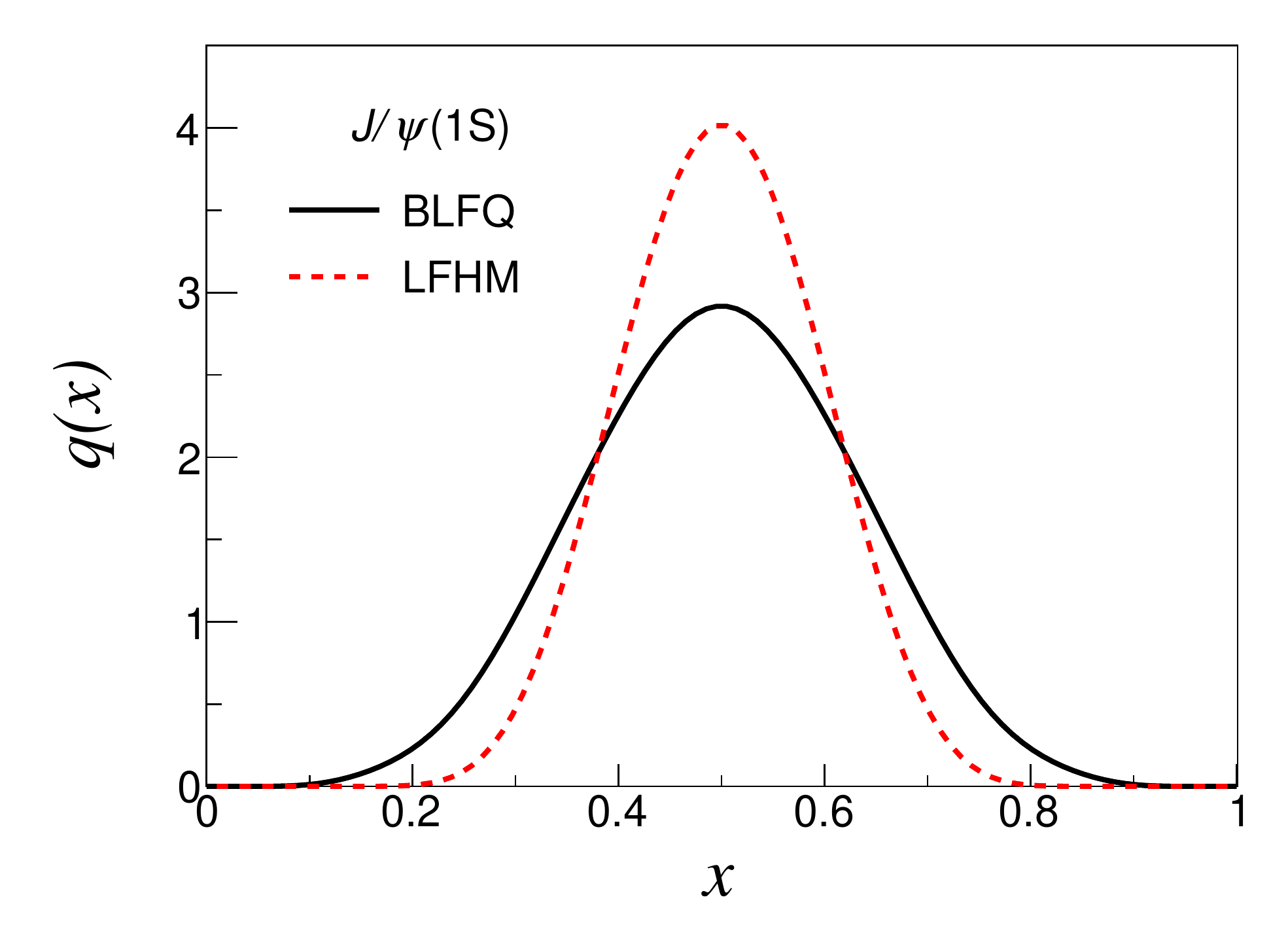}
\includegraphics[width=0.45\textwidth]{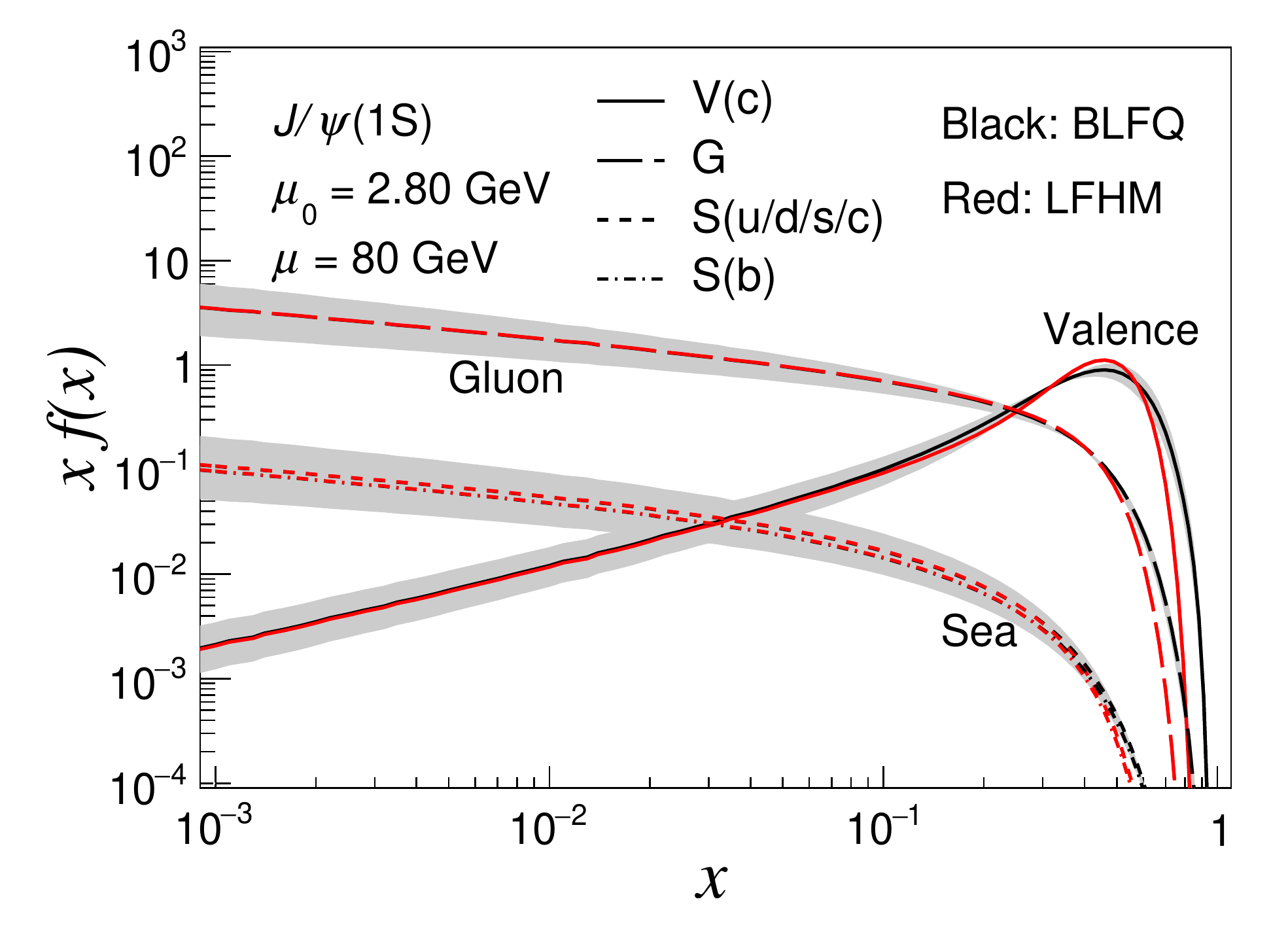}
\includegraphics[width=0.45\textwidth]{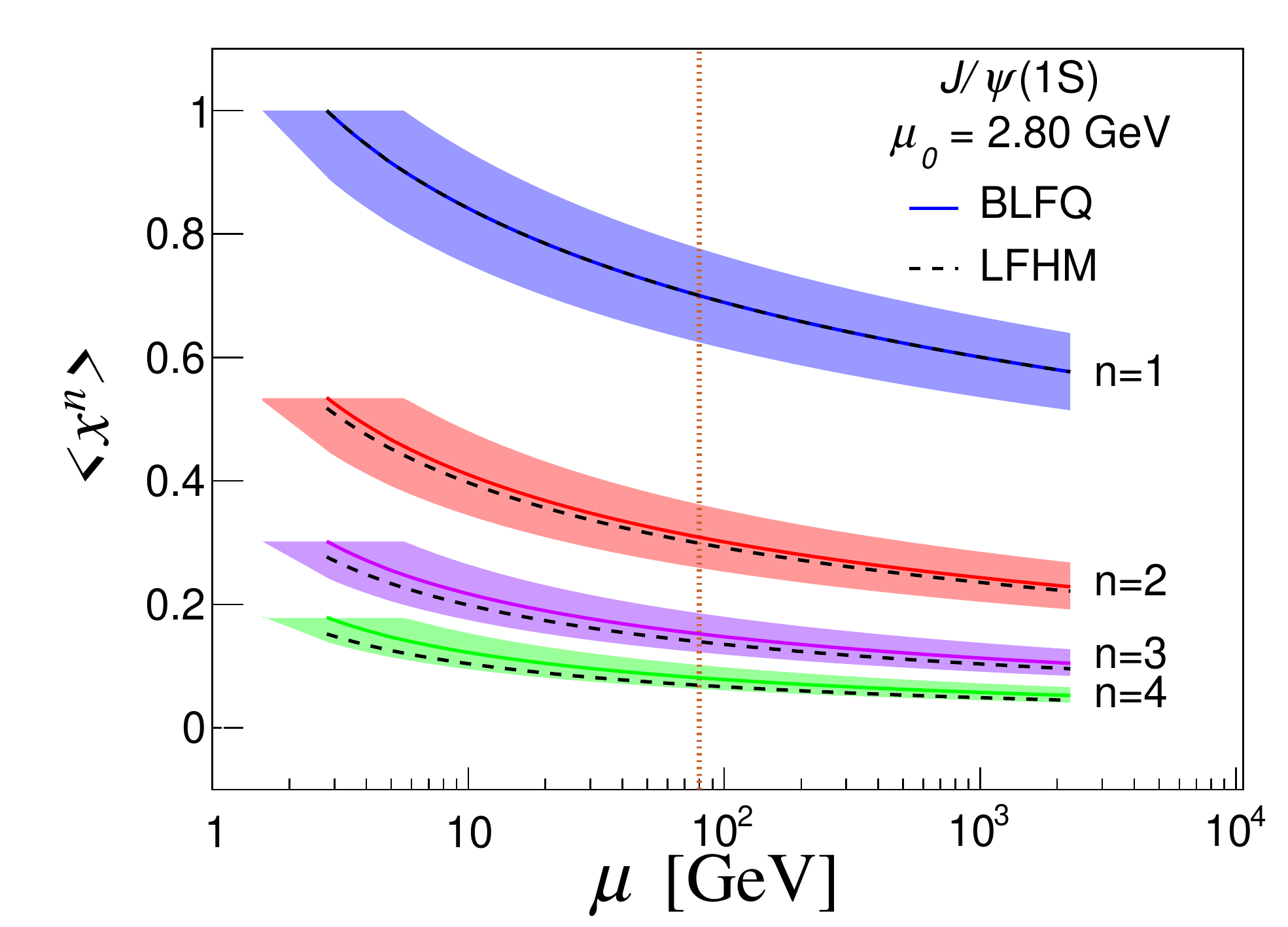} 
\caption{Comparison of the $J/\psi(1\rm S)$ PDFs within our BLFQ and the LFHM~\cite{Vega:2009zb}.
}
\label{fig_BLFQ_LFHQCD}
\end{center}
\end{figure*}

\begin{figure*}[htp]
\begin{center}
\includegraphics[width=0.85\textwidth]{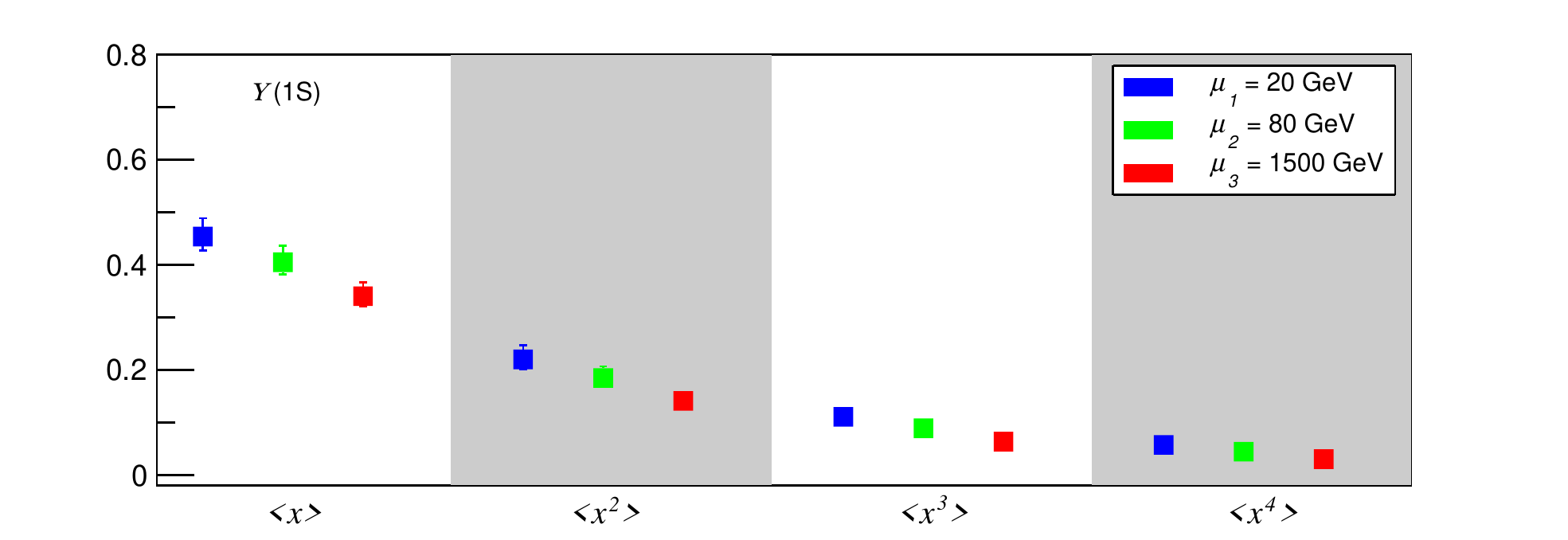}
\includegraphics[width=0.85\textwidth]{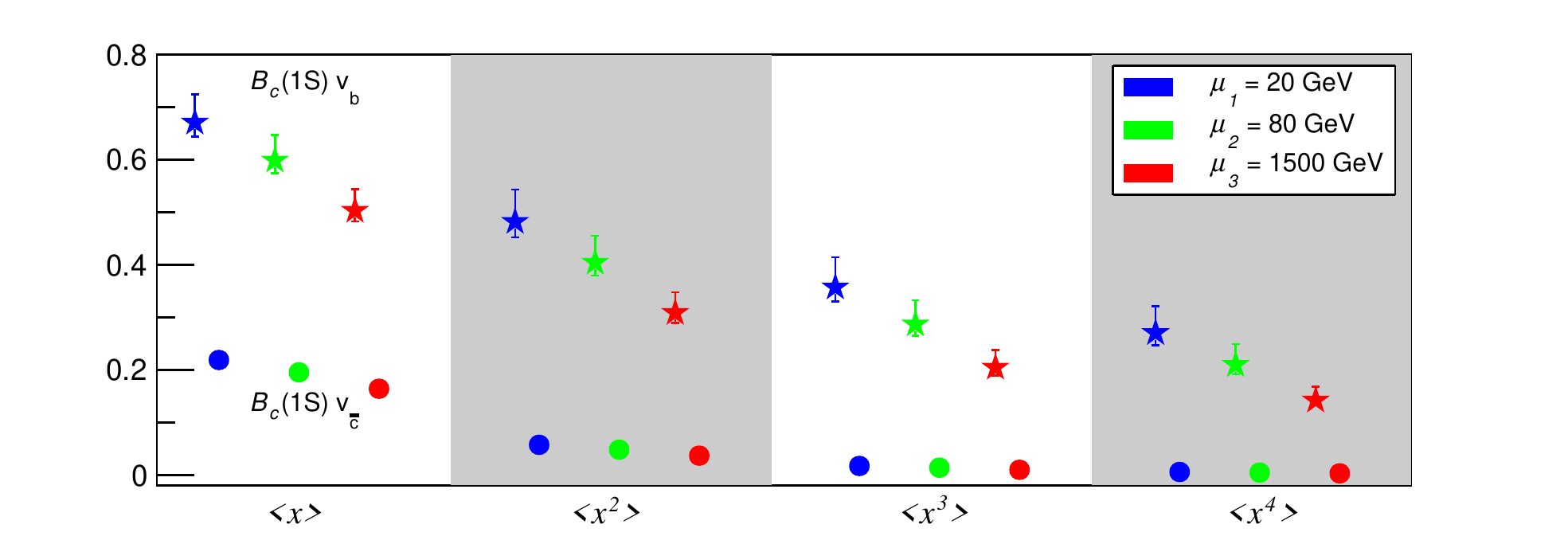}
\caption{The lowest four moments of valence quark distribution in $\Upsilon$ (1S) and $B_c$ (1S) at different scales. The solid markers with bars are the results of the present work taking into account the uncertainty in the initial scale ${\mu_{0}}$. The markers with different color correspond to the different final scales: $\mu_1=20$ GeV (blue), $\mu_2=80$ GeV (green), and $\mu_3=1500$ GeV (red).
}\label{fig_vxn}
\end{center}
\end{figure*}

To further explore the basis truncation effect, we compare the PDFs calculated from the leading basis function that excludes the one-gluon exchange effects and is reminiscent of light-front holography~\cite{Vega:2009zb}. We present a comparative study of the BLFQ results with our simulation in a light-front holography-inspired model  results for the $J/\psi (\rm 1S)$ PDF. We compare the initial scale PDFs within our BLFQ and the light-front holographic model (LFHM) in the left panel of Fig.~\ref{fig_BLFQ_LFHQCD}, whereas in the right panel of Fig.~\ref{fig_BLFQ_LFHQCD}, the evolved PDFs at the scale relevant to eRHIC, 80 GeV, are compared. In the LFHM, the valence quark and antiquark together carry the entire light-front momentum of the $J/\psi (\rm 1S)$ at the initial scale. Thus, we use the same initial scale as  BLFQ for our simulation of the LFHM.  We observe that our initial scale PDF is wider than that in the LFHM.
Meanwhile, the evolved PDFs exhibit good agreement at low $x$, however, there is a discrepancy between these two models at large $x$. For further investigation, we evaluate the four lowest nontrivial moments of the valence quark PDFs for the $J/\psi$(1S). In Fig. ~\ref{fig_BLFQ_LFHQCD} (lower panel), we show these results as a function of $\mu$ and compare with our simulation of the LFHM for the $J/\psi$(1S). The results are in good agreement. The four lowest nontrivial moments of $\Upsilon$(1S),
and $B_c$(1S) at different scales are shown in Fig.~\ref{fig_vxn}.

\begin{figure*}[htp]
\begin{center}
\includegraphics[width=0.47\textwidth]{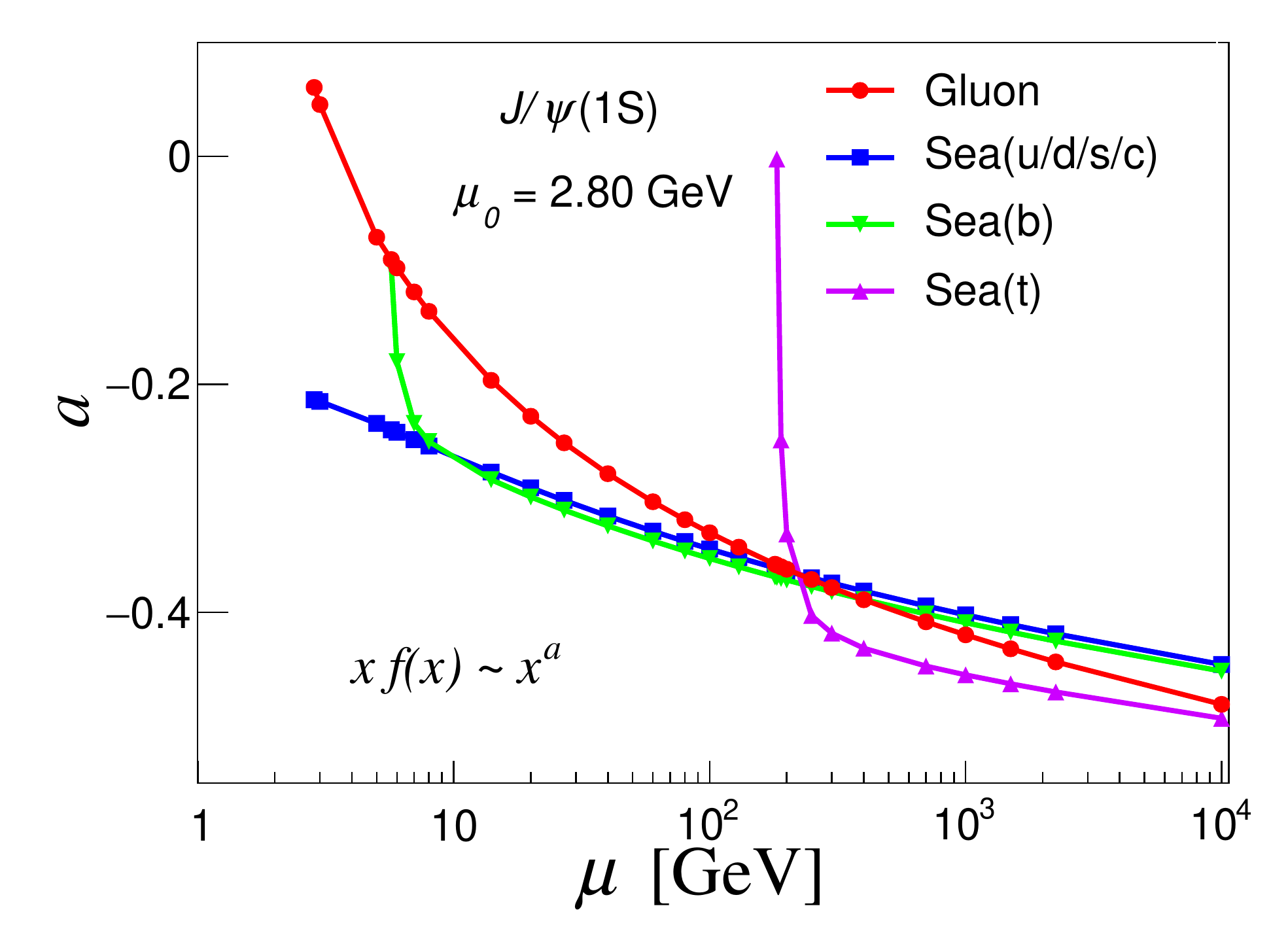}
\caption{$a$ as a function of $\mu$. At a low $x$ ($0.001<x<0.1$), the $x$-PDFs behave as $x f(x)\sim x^a$. The initial scale of the $J/\psi(1\rm S)$ PDFs for the basis truncations $N_{\rm max}=8$ is $\mu_0=2.80$ GeV. The red, blue, green, and magenta lines represent gluon, sea quark ($u/d/s/c$), sea quark ($b$), and sea quark ($t$), respectively.}
\label{fig_slope}
\end{center}
\end{figure*}

It is interesting to note that at low $x$, the $x$-PDFs behave like $x^a$ where $a>0$ for the valence quark, while for the sea and the gluon, $a>-1$. With the increasing  scale $\mu$, $a$ decreases and at the limit, $\mu\rightarrow \infty$, $a\rightarrow 0$ for the valence quark, and for the sea quark and the gluon, $a\rightarrow -1$. This phenomenon is independent of the PDFs at the initial scale. To demonstrate the low $x$ behavior of the gluon and the sea quarks PDFs with increasing scales, we consider $J/\psi(1\rm S)$ $x$-PDFs at low $x$, $xf(x)\sim x^a$ and show  the behavior of $a$ as a function of $\mu$ in Fig. \ref{fig_slope}. We notice that with increasing scale, $a$ falls, steadily, faster for the gluon than that for the sea quarks. This phenomenon again implies that the gluon dominates the distribution at low $x$ as the scale increases.

\section{summary}\label{summary}
We presented a comprehensive  study of the PDFs using the wave functions of a light-front model for quarkonium that incorporates light-front holography and the one-gluon exchange interaction with running coupling. The LFWFs have been obtained by using the BLFQ approach. We presented the results for the PDFs of $\eta_c(1\rm S)$, $J/\psi(1\rm S)$, $\psi(2\rm S)$, $\chi_{c2}(1\rm P)$ (charmonium),  $\eta_b(1\rm S)$, $\Upsilon(1\rm S)$, $\Upsilon(2\rm S)$, $\chi_{b2}(1\rm P)$ (bottomonium), and $B_c (\rm 1S)$, $B_c(\rm 2S)$, $B_c(\rm 1P)$ ($B_c$ meson) states. We observed that the qualitative behavior of charmonium $1\rm S~ (\eta_c~\rm{and}~J/\psi)$, $2\rm S~(\psi)$, and $1\rm P~(\chi_{c2})$ states is similar to their corresponding bottomonium $1\rm S~ (\eta_b~\rm{and}~\Upsilon)$, $2\rm S~ (\Upsilon)$ and $1\rm P~ (\chi_{b2})$ states. But, due to the smaller masses of charmonium compared to bottomonium, the width of the distributions is larger for charmonium. For bottomonium and charmonium, the PDFs are symmetric about $x=0.5$, while the peak moves to a different $x$ region following the naive mass fraction of the respective constituent quark mass to the total meson mass for the unequal constituent masses in $B_c$.

The QCD scale evolution of the heavy quark PDFs, which provides us with the knowledge of the gluon and the sea quark distributions, has also been investigated. We found that although the valence quark dominates at the large $x(>0.1)$ region, at the small $x$ region the distributions are mainly dominated by the gluon distribution. The momenta carried by the sea quark and gluon increase with increasing scale $\mu$. We observed that there is some sort of universality of the gluon PDFs from different states. Overall, the QCD evolution of heavy quark PDFs provides predictions for a wealth of information of the gluon and the sea quarks appearing in higher Fock sectors. For further improvement, future developments should focus on the inclusion of higher Fock sectors to explicitly incorporate sea quark and gluon degrees of freedom at appropriate initial scales. Our work provides a prediction of the expected data for heavy quarkonia PDFs from the future experiments as well as a guidance for the theoretical investigations of the PDFs with higher Fock components.

\section{Acknowledgments}
We thank S. Jia and N. Xu for many useful discussions. C.M. is supported by the National Natural Science Foundation of China (NSFC) under the under the Grants No. 11850410436 and No. 11950410753.  C. M. is also supported by the new faculty startup funding by the
Institute of Modern Physics, Chinese Academy of Sciences. This work of J. L., C. M., and X. Z. is also
supported by the Strategic Priority Research Program of the Chinese Academy of Sciences, Grant No. XDB34000000. M. Li is supported in part by the European Research Council (ERC) under the European Union’s Horizon 2020 research and innovation programme 
(Grant Agreement No. ERC-2015-CoG-681707). The content of this article does not 
reflect the official opinion of the European Union and responsibility for the 
information and views expressed therein lies entirely with the authors.
 This work of X.Z. is supported by new faculty
startup funding by the Institute of Modern Physics, Chinese Academy of Sciences and by Key Research Program of Frontier Sciences, CAS, Grant No ZDBS-LY-7020. J.P.V. is supported in part by the Department of Energy under Grants No. DE-FG02-87ER40371 and No. DESC00018223 (SciDAC-4/NUCLEI). A portion of the computational resources were provided by the National Energy Research Scientific Computing Center (NERSC), which is supported by the Office of Science of the U.S. Department of Energy under Contract No. DE-AC02-05CH11231.


\end{document}